\documentclass[twocolumn]{aastex63}
\usepackage{comment}
 \usepackage{hyperref}
\usepackage{amsmath}
 \usepackage{xfrac}
\usepackage{xcolor}
 \usepackage{ulem}

\shorttitle{Two-photon production}
 \shortauthors{Kulkarni \&\ Shull}
 
\graphicspath{{./}{Figures/}}

\newcommand {\HI}    {\ion{H}{1}}  %  HI
\newcommand {\HII}   {\ion{H}{2}}  %  HII
\newcommand{\Lya}{Ly$\alpha$}

\newcommand{\Ha}{\mbox{H$\alpha$}}

\begin{document}

\title{Two-photon production in low-velocity shocks}

\correspondingauthor{S.\ R.\ Kulkarni}
 \email{srk@astro.caltech.edu}

\author[0000-0001-5390-8563]{S.\ R.\ Kulkarni}
\affiliation{Owens Valley Radio Observatory 249-17, California
Institute of Technology, Pasadena, CA 91125, USA}

\author{J.\ Michael Shull}
 \affiliation{CASA, Department of Astrophysical \&\ Planetary Sciences, 
 University of Colorado, Boulder, CO 80303}
\affiliation{Department of Physics \&\ Astronomy, University of North Carolina, 
Chapel Hill, NC 27599 USA}

 \begin{abstract}
The Galactic interstellar medium  abounds in low-velocity shocks
with velocities $v_s\lesssim 70\,{\rm km\,s^{-1}}$. Some are
descendants of higher velocity shocks, while others start off at
low velocity (e.g., stellar bow shocks, intermediate velocity clouds,
spiral density waves).  Low-velocity shocks cool primarily via
Ly$\alpha$, two-photon continuum, optical recombination lines (e.g.,
H$\alpha$), free-bound emission, free-free emission and forbidden lines
of metals.  The dark far-ultraviolet (FUV) sky, aided by the fact
that the two-photon continuum peaks at 1400\,\AA, makes the FUV band 
an ideal tracer of low-velocity shocks. Recent {\it GALEX} FUV images
reaffirm this expectation, discovering faint and large interstellar 
structure in old supernova remnants and thin arcs stretching across 
the sky. Interstellar bow shocks are expected from fast stars from 
the Galactic disk passing through the numerous gas clouds in the 
local interstellar medium within 15 pc of the Sun. Using the best atomic
data available to date, we present convenient fitting formulae for
yields of Ly$\alpha$, two-photon continuum and H$\alpha$ for pure
hydrogen plasma in the temperature range of $10^4\,$K to $10^5\,$K.
The formulae presented here can be readily incorporated into
time-dependent cooling models as well as collisional ionization 
equilibrium models.

\vspace{1cm}
 \end{abstract}

%\keywords{instrumentation: photometric and spectrographs --- surveys
 %--- supernovae: general --- catalogs}

\section{Motivation}
 \label{sec:Motivation}

Supernova remnants and stellar wind bubbles are iconic examples of
shocks in the interstellar medium (ISM). These shocks, with the
passage of time, descend to lower velocities. Our interest here is
shocks with velocities less than $70\, {\rm km\,s^{-1}}$.  The
post-shock temperature depends on the mean molecular mass, but we
adopt a fiducial value of $T_s\le 10^5$\,K and investigate the
cooling of such shock-heated hydrogen gas.  These shocks cool 
primarily via Ly$\alpha$ (whose photons are trapped within the 
shocked region and eventually die on a dust particle) and two-photon
continuum.  The latter can be detected by Far Ultra-Violet (FUV)
imagers.  Low-velocity shocks
can also arise on Galactic length scales: intermediate-velocity and
high-velocity clouds raining down from the lower halo into the disk
and gas that is shocked as it enters a spiral arm.  \citet{V17}
provides a good description of the Milky Way's spiral arms, and
\citet{kko08} discuss Galactic interstellar shocks.

Stellar bow shocks are another major source of low-velocity shocks.
For instance, consider our own Sun, a generic G5V star with a weak
stellar wind ($2\times 10^{-14}\,M_\odot\,{\rm yr^{-1}}$) moving
into a warm ($\sim  7,000\,$K) and partially ionized cloud (ionization
fraction, $x\approx 1/3$) at a relative speed of 23--26\,km\,s$^{-1}$
\citep{frs11, mab+12,zhw+13, gj14}.  Because this velocity is not
larger than the magnetosonic velocity of the interstellar cloud,
there is only a ``bow wake" instead of a bow shock \citep{mab+12}.
In the Galactic disk, interstellar space is occupied by the Warm
Neutral Medium (WNM; $10^3$\,K to $8\times 10^3$\,K), the Warm
Ionized Medium (WIM; $8\times 10^3$\,K), and the Hot Ionized Medium
(HIM; $10^5$\,K to $10^6$\,K), in roughly equal proportions.  

From studies with SDSS-Apogee + {\it Gaia}-DR2 \citep{amh+20}, the 
3D velocity dispersion of the typical ($\alpha$-abundance tagged) 
thin-disk star is 48~km\,s$^{-1}$, whereas those belonging to the 
thick disk have dispersion of 87\,km\,s$^{-1}$.  The majority of 
these local stars reside in the thin disk with a density ratio
$n_{\rm thin} / n_{\rm thick} = 2.1 \pm 0.2$.  As discussed in 
a previous study (Shull \& Kulkarni 2023), a sizeable number of 
stars should be moving supersonically through ambient gas in the
WNM and WIM.\footnote{Only a few stars are likely transiting the
Cold Neutral Medium (CNM; 100\,K), given its small volume filling 
factor, $\sim 1\%$.} The sizes of the resulting bow shocks will be 
determined by the stellar velocity and the magnitude 
of the stellar wind.

Separately, recent developments warrant a closer look at low-velocity shocks.
We draw attention to the discoveries of three large-diameter supernova
remnants \citep{fdw+21} and a 30-degree long, thin arc in Ursa Major
\citep{bba+20}. In large part, these findings were made possible
with a new diagnostic -- {\it GALEX} FUV continuum imaging.  The
detection of such faint, extended features demonstrates simultaneously
the value of the dark FUV sky \citep{C87} as well as the
value of the FUV
band in detecting two-photon emission, a distinct diagnostic
of warm ($T\lesssim 10^5\,$K) shocked gas \citep{K22}.  

The primary goal of this paper is to develop accurate hydrogen
plasma cooling models, paying attention to the production of the
two-photon continuum in warm plasma, $T\lesssim 10^5\,$K, the
temperature range of interest to low velocity shocks. To this end,
we first derive the probability of Ly$\alpha$, two-photon continuum,
and H$\alpha$ resulting from excitation of the ground state of
hydrogen to all $n\ell$ levels for $n\le 5$
(\S\ref{sec:TwophotonProduction}).  Next, we review rate coefficients
for line excitation by collisions with electrons
(\S\ref{sec:LineEmission}), followed by a review of collisional
ionization (\S\ref{sec:Collisional_Ionization}). The results 
are combined to construct a cooling curve for warm hydrogen plasma
(\S\ref{sec:HydrogenCoolingCurve}).  We then present a comprehensive
(isobaric and isochoric) cooling framework and apply it to gas
shock heated to $10^5\,$K (\S\ref{sec:SimpleCoolingModel}).
In \S\ref{sec:ConclusionProspects} we summarize our results and
discuss future prospects.  Unless otherwise stated, the atomic line
data (A-coefficients, term values) were obtained from the NIST Atomic Spectra
Database\footnote{\url{https://physics.nist.gov/PhysRefData/ASD/lines_form.html}}
and basic formulae are from \citet{D11}.

\begin{comment}

\begin{figure}[htbp]
 \plotone{4p.png}
  \caption{\small Ladder for two-photon decay following excitation
  to any of the $n=4$ levels of hydrogen (energy separation not to
  scale).  As explained in the text, we assume that case B holds.
  Therefore, we do not display Lyman transitions ($n$p-1s). As can
  be see from the figure, transitions to 3p or 2s lead to two-photon
  decay.}
 \label{fig:4p_ladder}
\end{figure}

\vspace{1cm}
\end{comment}

\begin{deluxetable}{llrr}[htb]
\tablecaption{The indexing scheme and spectroscopic terms}
\tablewidth{0pt}
\tablehead{\colhead{$i$} & \colhead{level} &
\colhead{$L_k$\,(cm$^{-1}$)} & \colhead{$k$}}
\startdata
1 & 1s & 0 & -\\
2 & 2s & 82303 & 1\\
3 & 2p & " & 2\\
4 & 3s & 97544 & 3\\
5 & 3p & " & 4\\
6 & 3d & " & 5\\
7 & 4s & 102879 & 6\\
8 & 4p & " & 7\\
9 & 4d & " & 8\\
10 & 4f & " & 9\\
11 & 5s & 105348 & 10\\
12 & 5p & " & 11\\
13 & 5d & " & 12\\
14 & 5f & " & 13\\
15 & 5g & " & 14\\
\enddata
 \tablecomments{In constructing this table we follow the notation
 and term values of \citet{abb+00}, where $i$ is the index assigned
 to levels, and $k$ is the index for upper levels excited in
 transitions from the ground state (1s). The energy for transition
 $k$ is $hcL_k$ where $L_k$ is the wavenumber (in cm$^{-1}$).  The
 symbol `"' is equivalent to ``ditto".  As can be gathered from the
 entries for $L_k$, the small  differences in energy due to fine
 structure effects are ignored.} 
	\label{tab:indexing}
\end{deluxetable}

\section{Two-photon production}
 \label{sec:TwophotonProduction}

Colliding electrons excite hydrogen atoms to various levels and,
if sufficiently energetic, ionize \HI\ to \HII.  Excited levels are
also populated by radiative recombination.  Excited hydrogen
atoms return to the ground state, some by emitting a Lyman-series
photon and others via a cascade of optical/IR recombination lines
and ending with Ly$\alpha$ emission. Atoms that find themselves in
the metastable {\it 2s}\ $^2{\rm S}_{1/2}$ level, if undisturbed over a 
timescale of $A_{2s\rightarrow 1s}^{-1}\approx 0.12\,$s, return to 
the ground state by emitting a two-photon continuum.  Here, $
A_{2s\rightarrow 1s}$ is the Einstein A-coefficient for the 
{\it2s}-{\it 1s} transition \citep{D86}.  Its value should be compared 
to those for allowed transitions (e.g., $6.26\times 10^8\, {\rm s}^{-1}$ 
for \Lya\ and$(1-5)\times 10^7\,{\rm s^{-1}}$ for H$\alpha$, depending 
on the upper levels, {\it 3s, 3p, 3d}, involved.)

The goal of this section is to compute the production of Ly$\alpha$
photons, two-photon continuum and H$\alpha$ resulting from electronic
excitation of H atoms. We consider excitations to 15 $n \ell$ levels;
see Table~\ref{tab:indexing} for term values and index scheme.
We make the following assumptions. (1) The proton  density
in the plasma is less than the ``{\it 2s} critical density" of 
$1.5\times10^4\,{\rm cm^{-3}}$  (see Chapter 14 of \citealt{D11}).  
This ensures that atoms in the {\it 2s} level are not collisionally mixed 
to the {\it 2p} level over a timescale of $A_{2s\rightarrow 1s}^{-1}$ and 
thus relax by emitting a two-photon continuum. (2) The cooling plasma is
optically thick to Lyman lines (case B), so that Lyman photons are
absorbed in the vicinity of where they are emitted. Thus, when
computing branching ratios, all allowed Lyman series recombinations
can be ignored.

\subsection{Photon Yields}

Consider, for example, an atom excited to one of the $n=3$ levels.
An atom excited to {\it 3s} or {\it 3d} will decay to {\it 2p} by 
emitting \Ha\ followed by \Lya.  (We ignore forbidden transitions such as 
{\it ns-1s} 
two-photon decays; see \citealt{cs08}.) An atom excited to {\it 3p} can 
decay by emitting Ly$\beta$ or decay to {\it 2s} by emitting \Ha\ 
followed by two-photon decay. For the latter, the branching fraction
$\mathcal{B}_\beta$ for Ly$\beta$ emission is $A_{3p\rightarrow
1s}/(A_{3p\rightarrow 1s}+A_{3p\rightarrow 2s}) \approx 88\%$.
However, under case~B, the Ly$\beta$ photon will be absorbed elsewhere
in the nebula, and the situation will be repeated until de-excitation
ends with emission of H$\alpha$+Ly$\alpha$.

\begin{deluxetable}{lrrrrr}[hbt]
\tablecaption{Lyman lines: Optical depths and scatterings}
\tablewidth{0pt}
\tablehead{\colhead{line} & \colhead{$\lambda$ (\AA)} &
\colhead{$f$} & \colhead{$\tau_0$} & 
\colhead{$n \ell$} &
\colhead{$\mathcal{B}$}} 
\startdata
Ly$\alpha$ & 1215.67 & 0.4164 & 1000 & {\it 2p} & 1 \\
Ly$\beta$ & 1025.73 & 0.07912 & 160 & {\it 3p} & 0.881 \\
Ly$\gamma$ & 972.54 & 0.02901 & 56&  {\it 4p} & 0.839\\
Ly$\delta$ & 949.74 & 0.01394 & 26 & {\it 5p} & 0.819\\
%Ly$\epsilon$ & 937.80 & 0.007799 \\
%Ly$\zeta$ & 930.74 & 0.004184 \\
%Ly$\eta$ & 926.22 & 0.003183 \\
%Ly$\theta$ & 923.15 & 0.002216 \\
%Ly$\iota$ & 920.96 & 0.001605 \\
\enddata
 \tablecomments{Columns 1--4 give the name, wavelength, absorption
 oscillator strength, and central optical depth of the line. The
 column density of the nebula is assumed to provide a line-center
 optical depth of $\tau_0=1000$ for \Lya, from which $\tau_0$ for
 other Lyman lines follow.  $\mathcal{B}$ (column 6) is the branching ratio
 for an atom excited to an {\it np} level (column 5) to relax by emitting the
 appropriate Lyman series line, as opposed to a multi-decay cascade.}
 \label{tab:LymanLines}
\end{deluxetable}

For a fiducial value of optical depth ($\tau_{0,\alpha} = 1000$)
of Ly$\alpha$, Table~\ref{tab:LymanLines} lists the corresponding
optical depths for the Lyman series.  The branching ratio 
$\mathcal{B}_\gamma$ to emit a Ly$\gamma$ line is slightly smaller 
that that for Ly$\beta$. As with Ly$\beta$ under case-B conditions, 
Ly$\gamma$ will also be converted to some combination of Ly$\alpha$, 
optical/IR recombination lines, and a two-photon continuum. The 
oscillator strength, $f\propto n^{-3}$,
where $n$ is the principal quantum number of the excited state.
Thus the Lyman-line optical depths decrease rapidly with increasing
$n$ (up the series).  In contrast, the branching factors $\mathcal{B}$
decrease slowly with $n$.

Each state other than {\it 4s} and {\it 5s} has two fine-structure levels.  
For example, the {\it 4p} state has two levels, $P_{1/2}$ and $P_{3/2}$, 
with very little energy difference between the fine structure levels.  
However, the electron collisional excitation rate coefficients presented 
below (\S\ref{sec:LineEmission}) refer to the sum of transitions to the 
entire level, e.g., {\it 1s}$\rightarrow${\it 4p}. 
The excitation coefficient is divided in proportion to the number of levels
of the excited state, $g_u=2J+1$ where $J$ is the total angular momentum
of the excited state.  The photon yields for  Ly$\alpha$, 2$\gamma$
continuum, and H$\alpha$ are given in  Table~\ref{tab:PhotonYields}.

\begin{deluxetable}{llrr}[htb] %PhotYield.m (run Branching.m, BR_levels.m first)
\tablecaption{Photon yields for Ly$\alpha$, H$\alpha$, and $2\gamma$ continuum}
\tablewidth{0pt}
\tablehead{\colhead{$k$} &\colhead{$p_k({\rm Ly}\alpha)$} 
&\colhead{$p_k({\rm H}\alpha)$} & \colhead{$p_{k}(2\gamma)$}}
\startdata
1 &  0 & 0 & 1\\
2 &  1 & 0 & 0 \\
3 &  1 & 1 & 0\\
4 &  0 & 1 & 1\\
5 &  1 & 1 & 0\\
% k & Ly-alpha & H-alpha & 2-phot \\
 6 & 0.585 & 0.415 & 0.415\\
 7 & 0.261 & 0.261 & 0.739\\
 8 & 0.813 & 0.187 & 0.187\\
 9 & 1 & 1 & 0\\
10 & 0.513 & 0.378 & 0.487\\
11 & 0.305 & 0.265 & 0.695\\
12 & 0.687 & 0.267 & 0.313\\
13 & 0.936 & 0.702 & 0.064\\
14 & 1 & 1 & 0
\enddata
 \tablecomments{Photon production yields $p_k$ upon excitation to
 level with index $``k"$, under case B conditions; see
 Table~\ref{tab:indexing} for the definition of $k$. For instance,
 an H atom excited to {\it 3s} ($k=3$) relaxes by emitting one H$\alpha$
 photon and one Ly$\alpha$ photon.}
  \label{tab:PhotonYields}
\end{deluxetable}

\section{Electron Collisional Excitation}
 \label{sec:LineEmission}

The excitation of lines of hydrogen due to collisions with electrons
is a venerable topic in ISM studies. The classic review by \cite{dm72}
summarizes the atomic physics of the 1960s.  \cite{sw91} undertook
detailed calculations of the $n = 1\rightarrow 2$ excitations and
also provided an estimate for the cooling rate coefficient,
$\Lambda_{\rm HI}$.  Anderson et al.\ (2002\nocite{abb+02}; see
also \citealt{abb+00}) present close-coupling R-matrix calculations.
We adopt these rates since they offer improved accuracy over  previous
studies \citep{Swb+90}.  The \citet{abb+02} theoretical cross
sections were constructed with 15 physical energy states up to $n
= 5$ ({\it 1s} to {\it 5g}) supplemented by 24 pseudo-states described by
orbitals ($\overline{n}, \overline{\ell}$) with $\overline{n} =
6-9$ and $\overline{\ell} = 0-5$.

\citet{abb+02} present collision strengths, $\overline\Omega_{ij}$,
for excitation from levels $i$ to $j$, averaged over a Maxwellian
velocity distribution at electron temperatures, $E_T \equiv k_BT$,
ranging from 0.5--25~eV.  The collisional excitation rate coefficients
(in cm$^3$~s$^{-1}$) are then given by:
 \begin{eqnarray} q_{i\rightarrow j} &=&\frac{2\sqrt\pi\ \alpha
 a_0^2}{g_i}
    \sqrt{\frac{I_{\rm H}}{k_BT}} \, \overline\Omega_{ij}(T) \,
       \exp (-E_{ij}/kT) \cr \cr
  &=& \frac{8.629\times 10^{-6}}{g_i}
 \frac{\overline\Omega_{ij}}{\sqrt{T}} \exp (-E_{ij}/k_B T) \;  \,
  \label{eq:q_ij} ,
 \end{eqnarray}
where $a_0$ is the Bohr radius, $\alpha$ is the fine structure
constant, $g_i$ is the degeneracy of level $i$, and $E_{ij}$ is the
energy difference between level $i$ and $j$. Since we are only
interested in excitations from the ground state, we assume $g_i=2$
for {\it 1s} $(^2$S$_{1/2})$.

\begin{figure}[htp] 	%Anderson_OscillatorStrengths.m
 \centering
  \includegraphics[width=3in]{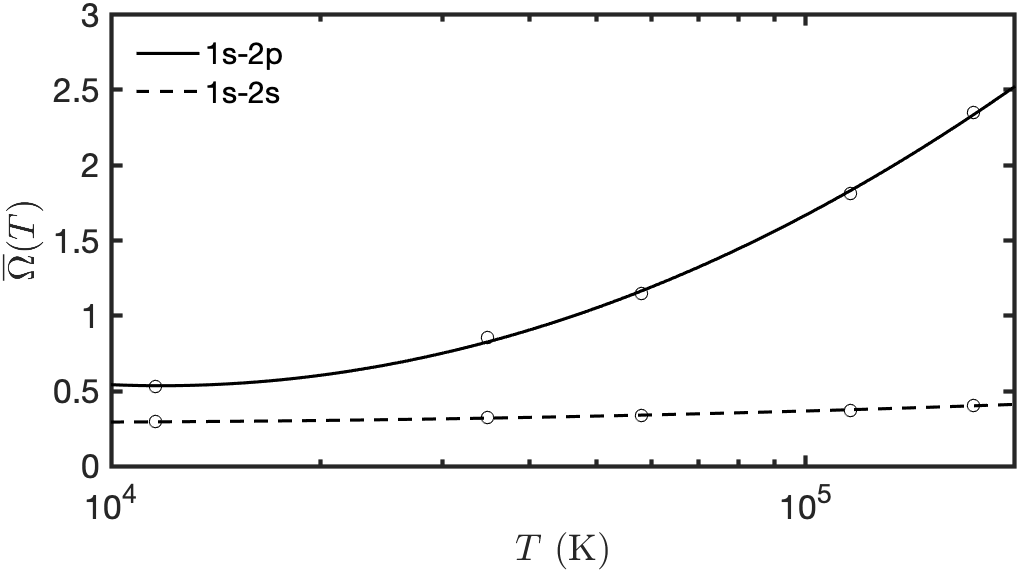}
   \includegraphics[width=3in]{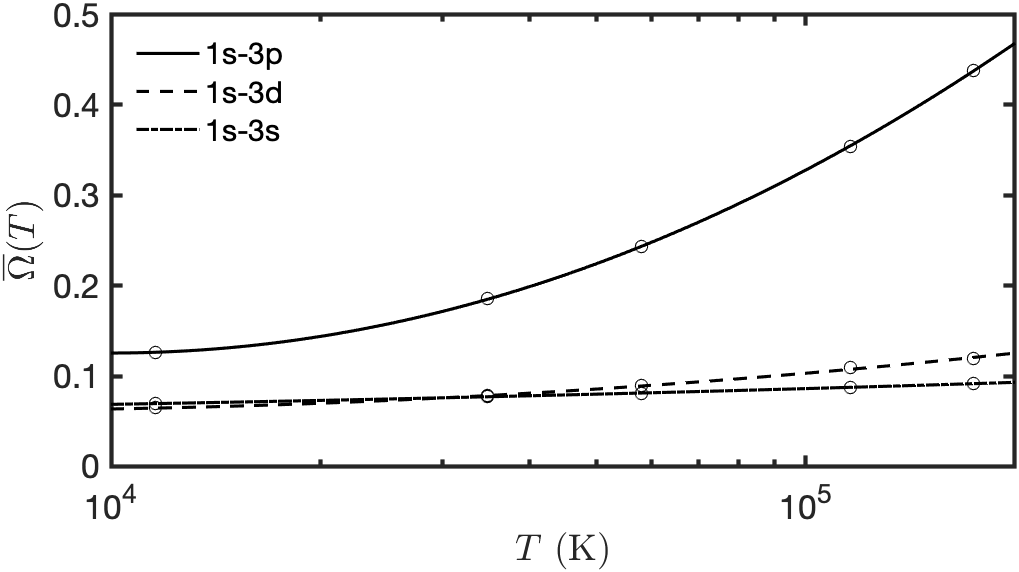}\\
    \includegraphics[width=3in]{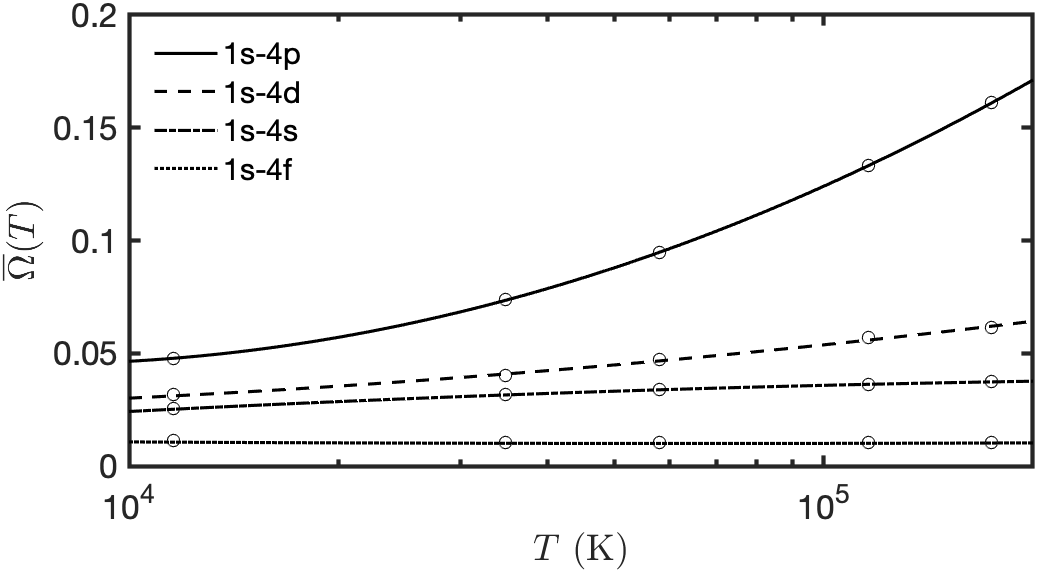}
     \includegraphics[width=3in]{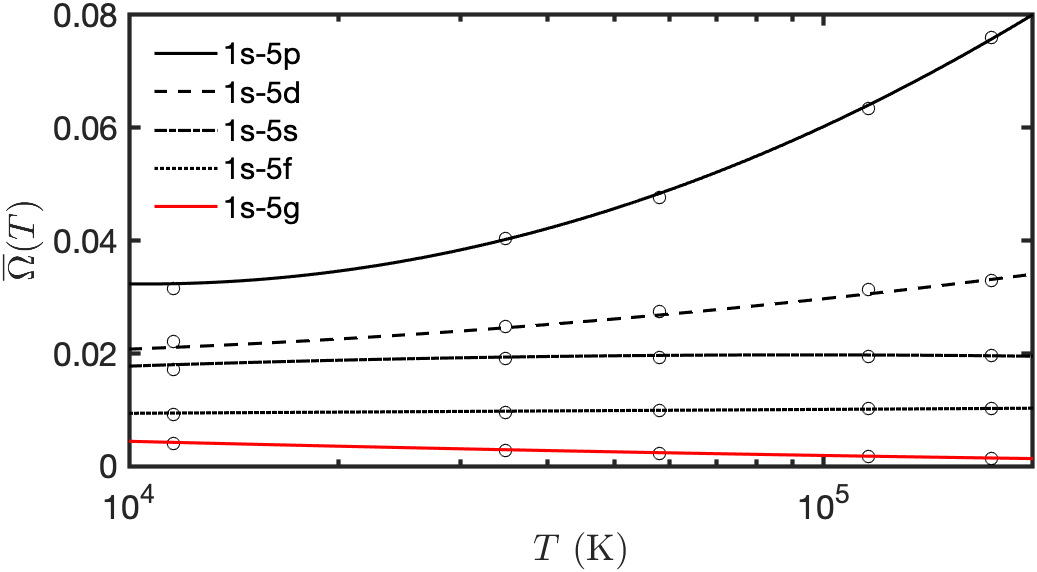}
      \caption{\small The electron-impact collision strengths,
      $\overline{\Omega}_{ij}$, for {\it 1s} to $(n,\ell)$ excitations
      of hydrogen as a function of temperature for $n=2,3,4,5$.
      Open circles are model data from \citet{abb+02}, and the 
      lines are second-order polynomial fits (see
      Equation~\ref{eq:model_function} and
      Table~\ref{tab:Anderson_polynomials}).}
 \label{fig:H2_H5}
\end{figure}

Given our focus on warm plasma, we limit the model fits 
to $1~{\rm eV} \le E_T \le 15\,$eV.  After some experimentation, we found 
that a cubic polynomial provides an adequate fit\footnote{A first-order 
fit would have been sufficient for excitations to all states but 
1s-np and 1s-nd.  For simplicity, we elected to use the
same number of coefficients for all transitions.}:
 \begin{equation}
	\overline{\Omega}_{ij}(T)=a_1+a_2x+a_3x^2 
		\label{eq:model_function} \; , 
 \end{equation}
where $x={\rm ln}(T/10^6\,{\rm K})$. The fit is precise to about
1\% for all levels except $5\ell$ levels, for which the fitting
errors approach 5\% (see Figure~\ref{fig:H2_H5}). The fitting
coefficients can be found in Table~\ref{tab:Anderson_polynomials}.

\begin{table}  %Anderson_OscillatorStrength.m
\caption{Polynomial fits to collision strengths\tablenotemark{a} } 
 \label{tab:Anderson_polynomials}
\begin{tabular}{llrrr}
\hline\hline
$k$ & trans & $a_0$ & $a_1$ & $a_2$ \\
\hline
 1 & 1s-2s & $0.5532$ & $0.1044$ & $0.0105$ \\
 2 & 1s-2p & $5.4261$ & $2.2029$ & $0.2481$ \\
 3 & 1s-3s & $0.1121$ & $0.0131$ & $0.0008$ \\
 4 & 1s-3p & $0.9355$ & $0.3518$ & $0.0382$ \\
 5 & 1s-3d & $0.1957$ & $0.0517$ & $0.0050$ \\
 6 & 1s-4s & $0.0390$ & $-0.0005$ & $-0.0008$ \\
 7 & 1s-4p & $0.3224$ & $0.1124$ & $0.0114$ \\
 8 & 1s-4d & $0.0944$ & $0.0213$ & $0.0016$ \\
 9 & 1s-4f & $0.0117$ & $0.0011$ & $0.0002$ \\
10 & 1s-5s & $0.0175$ & $-0.0019$ & $-0.0004$ \\
11 & 1s-5p & $0.1464$ & $0.0501$ & $0.0055$ \\
12 & 1s-5d & $0.0471$ & $0.0094$ & $0.0008$ \\
13 & 1s-5f & $0.0108$ & $0.0003$ & $-0.0000$ \\
14 & 1s-5g & $0.0005$ & $-0.0004$ & $0.0001$\\
\hline
%\hrulefill
\end{tabular}
\tablenotetext{a}{Coefficients ($a_i$) for Maxwellian-averaged 
collision strengths, fitted by 
$\overline{\Omega}_k = \sum_{i=0}^2 a_ix^i$,
where $x={\rm ln}(T/10^6\,{\rm K})$. Here, $k$ is the index 
of the transition (see Table~\ref{tab:indexing}).}
\end{table}

\begin{figure*}[htbp] %plot_Anderson_cooling_coefficients.m
 \centering
  \includegraphics[width=5.5in]{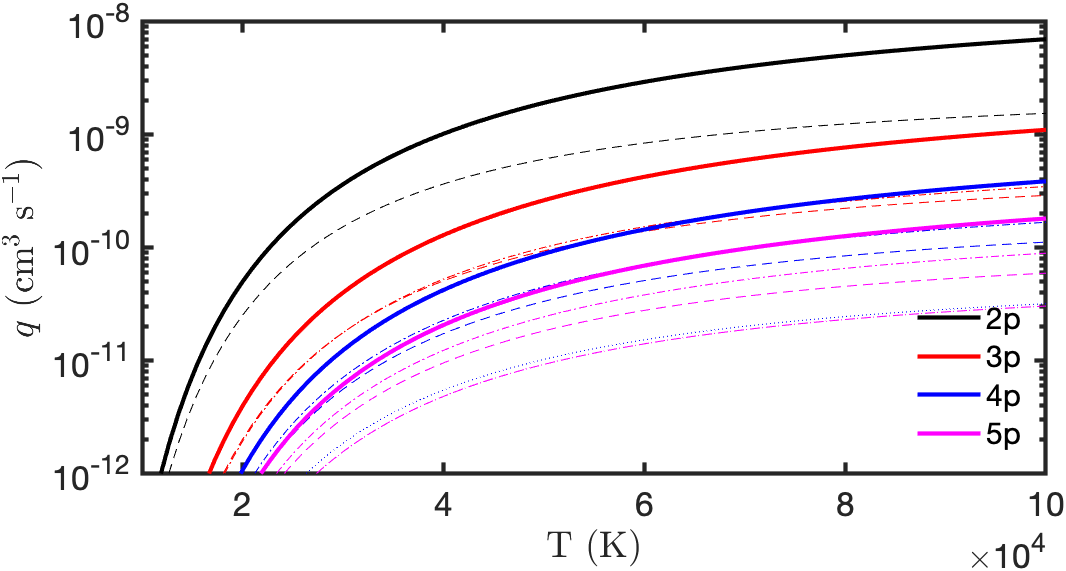}  
   \caption{\small Electron collisional excitation rate coefficients,
   $q_{ij}$,  for {\it 1s}$\rightarrow (n,\ell)$ transitions of hydrogen
   derived from collision strengths provided by \citet{abb+02}.
   The curves are coded by color ($n=2,3,4,5$ as labeled) and by
   line type (dash-dash 1s-$n$s, continuous for {\it 1s-np}, dash-dot
   for {\it 1s-nd}, dotted for {\it 1s-nf}, and back to dash-dash for
   {\it 1s-ng}.)  \vspace{0.5cm}  } 
 \label{fig:Anderson_q}
\end{figure*}

The line cooling rate per unit volume is given by $n_e n_{\rm
HI}\Lambda_{\rm HI}$ where $n_e = n_p$ is the electron (and proton)
density and $n_{\rm HI}$ is the density of H atoms.  The total
particle density is $n_t=n_{\rm HI}+n_e+n_p=n_{\rm H}(1+x)$ with
$n_{\rm H}=n_p+n_{\rm HI}$ and $x=n_e/n_{\rm H}$.

We used the fitting model to compute the run of collisional rate
coefficients, $q_{i\rightarrow j}$,  with temperature
(Figure~\ref{fig:Anderson_q}).  With the cooling coefficients in
hand, we computed the sum of the luminosity radiated in lines up
to $n=5$. We consider this sum to be an adequate representation of
$\Lambda_{\rm HI}(T)$ for warm hydrogen.  The cooling rate coefficient
is
 \begin{equation}
  \Lambda_{\rm HI}(T) = \sum_{k=1}^{14} q_k(T) E_k
   \label{eq:Lambda_HI_formal}  \; ,
 \end{equation}
where the energy of transition with index $k$ is $E_k=hcL_k$; see
Table~\ref{tab:indexing} for definition of $k$ and the adopted
values for the wavenumbers,  $L_k$ (in cm$^{-1}$).  Separately, in
\S\ref{sec:OtherCoolingModels}, we compare this cooling coefficient
to previously published coefficients \citep{S78,sw91,dlm+97}.

The coefficient for energy loss through line $X$ (where, for 
instance, $X$ denotes Ly$\alpha$, H$\alpha$, $2\gamma$) is given by
 \begin{equation*}
  \Lambda_X(T) = \sum_{k=1}^{14}q_{k}(T)p_k(X)E_X  \; ,
 \end{equation*}
where $E_X$ is line energy and $p_k(X)$ is given in
Table~\ref{tab:PhotonYields}.

\subsection{Simple Fits to Line cooling and Collision rates} 

\begin{figure}[htbp]    %Polyfit_Cooling_Collision.m
 \plotone{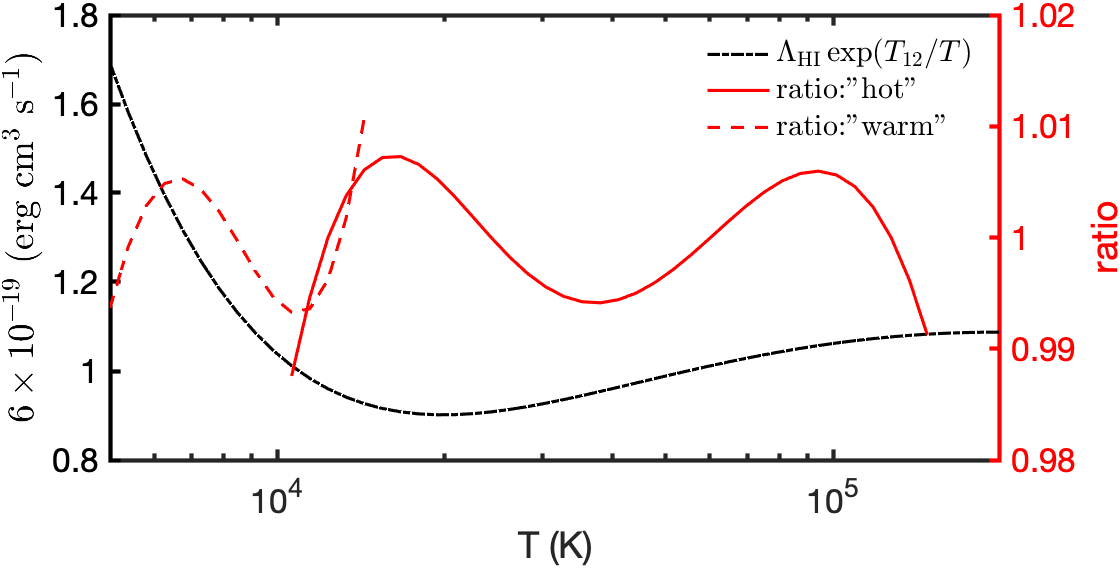}
  \plotone{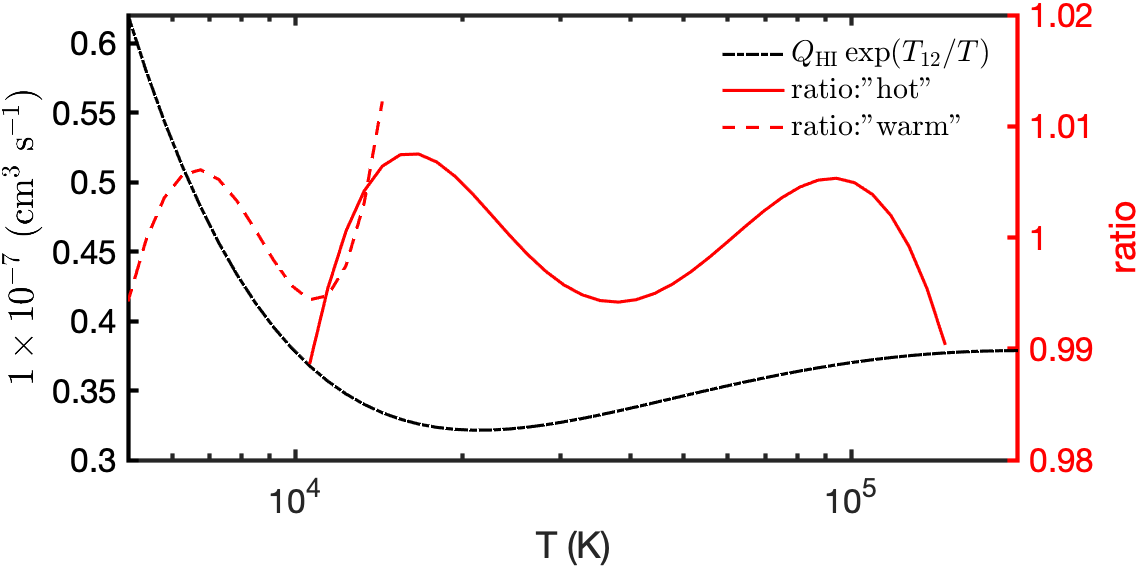}
   \caption{\small (Top Panel.)  Left axis shows the line cooling
   coefficient, $\Lambda_{\rm HI}(T)$, for hydrogen (black line),
   where the total cooling rate is $n_e n_{\rm HI} \Lambda_{\rm
   HI}(T)$.  Right axis shows the percent residuals (red-dashed
   lines) in the form of $[1- (\Lambda_{\rm HI}/{\rm fit})]$.  The
   model form is given by Equation~\ref{eq:Q}, and model parameters
   are given in Table~\ref{tab:Cooling_Collisional_Coefficients}.
   (Bottom Panel.) The same, but for $Q$, the collisional coefficient.
   The rate of collisions per unit volume is $n_en_{\rm HI}Q$.}
 \label{fig:LambdaHI_model_fit}
\end{figure}

The collisional excitation rate coefficient is the sum over all
hydrogen levels,
 \begin{equation*}
     Q(T) = \sum_{i=1}^{14}q_k(T)  \; .
 \end{equation*}
Both $Q(T)$ and the hydrogen cooling rate (from excitation to $n=2$)
fall off with temperature as $\exp(-T/T_{12})$, where $k_BT_{12}=3
I_{\rm H}/4$ is the energy difference between $n=1$ and $n=2$ levels.
We fit the collision rate and $\Lambda_{\rm HI}$ over two temperature
ranges: ``hot" ($10^4~{\rm K} < T <1.5\times 10^5\,$K) and ``warm"
($10^4~{\rm K} < T < 1.5\times 10^4~{\rm K}$),
 \begin{equation}
     Q_{\rm HI}(T) = A\exp(-T/T_{12})\sum_{i=0}^{n}a_i z^i \label{eq:Q}
     \; ,
 \end{equation}
where $z= \log T_4$ with $T_4 = (T/10^4\,{\rm K})$.  A similar expression 
was derived for $\Lambda_{\rm HI}(T)$.  The fitting parameters for
$Q_{\rm HI}$ and $\Lambda_{\rm HI}$ are given in
Table~\ref{tab:Cooling_Collisional_Coefficients}, and the quality
of the fit is displayed in Figure~\ref{fig:LambdaHI_model_fit}.

\begin{deluxetable}{rrrrrr}[hbt]
\tablecaption{Fits to Cooling and Collisional Coefficients}
 \label{tab:Cooling_Collisional_Coefficients}
 \tablewidth{0pt}
 \tablehead{
 \colhead{Quantity} & 
 \colhead{$A$} & 
 \colhead{$a_0$} &
 \colhead{$a_2$} &
  \colhead{$a_3$} &
 \colhead{$a_4$}
 }
\startdata
%y=a1+a2*x+a3*x^2+a4*x^3; x=log10(T/1e4)
%HI-cooling coefficients
$\Lambda_{\rm HI}$:hot & $6.0\times 10^{-19}$ & $1.018$ & $-0.771$ & $1.537$ & $-0.716$  \\
$\Lambda_{\rm HI}$:warm & $6.0\times 10^{-19}$ & $1.032$ & $-1.138$ & $3.376$  \\
\hline
% HI collisional-coefficient
$Q_{\rm HI}$:hot & $1.0\times 10^{-7}$ & $0.371$ & $-0.304$ & $0.560$ & $-0.255$  \\
$Q_{\rm HI}$:warm & $1.0\times 10^{-7}$ & $0.376$ & $-0.433$ & $1.220$  \\
\enddata
 \tablecomments{``Quantity" refers to the cooling coefficient 
 ($\Lambda_{\rm HI}$ in erg cm$^3$ s$^{-1}$) or collisional excitation 
 rate coefficient ($Q_{\rm HI}$ in cm$^3$~s$^{-1}$). These quantities
 are fitted to a model displayed in Equation~\ref{eq:Q} over two
 temperature ranges: ``hot" ($10^4~{\rm K} < T <1.5\times 10^5~{\rm
 K}$) and ``warm" ($10^4~{\rm K} < T < 1.5\times 10^4~{\rm K}$.)}
\end{deluxetable}

\begin{figure*}[htbp]   %plot_Hcooling_warm.m
 \centering
  \includegraphics[width=4.5in]{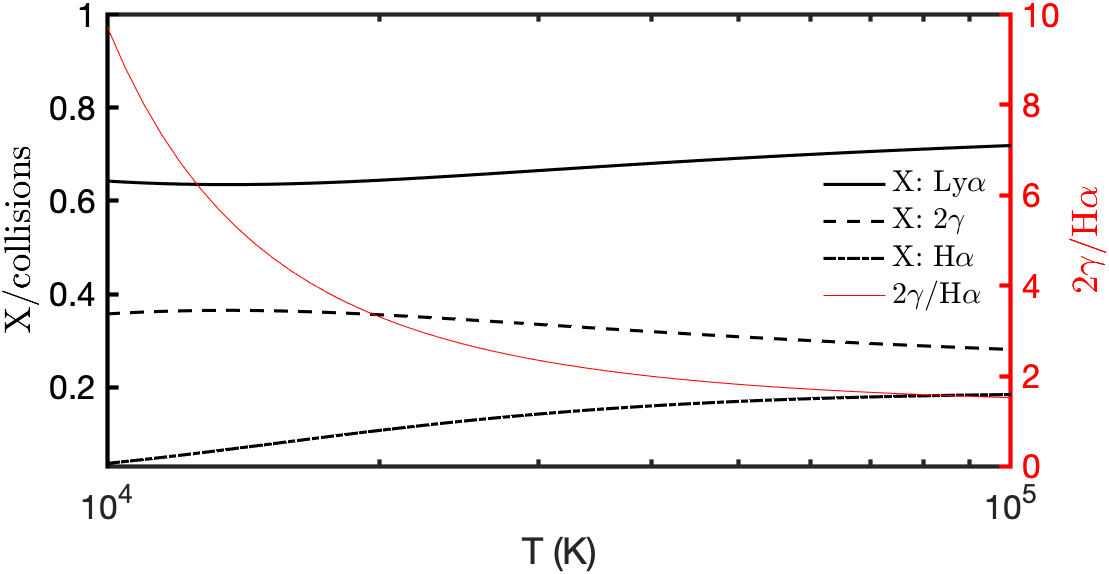}
   \caption{\small ({\it Left}) Case~B photon production per collision
   as a function of temperature, $T$ for Ly$\alpha$, H$\alpha$ and
   two-photon continuum.  ({\it Right}) The ratio of yields of
   two-photon decays to that of H$\alpha$ photons.}
 \vspace{1cm} 
  \label{fig:Warmcooling_photon}
\end{figure*}

\begin{figure}
 \plotone{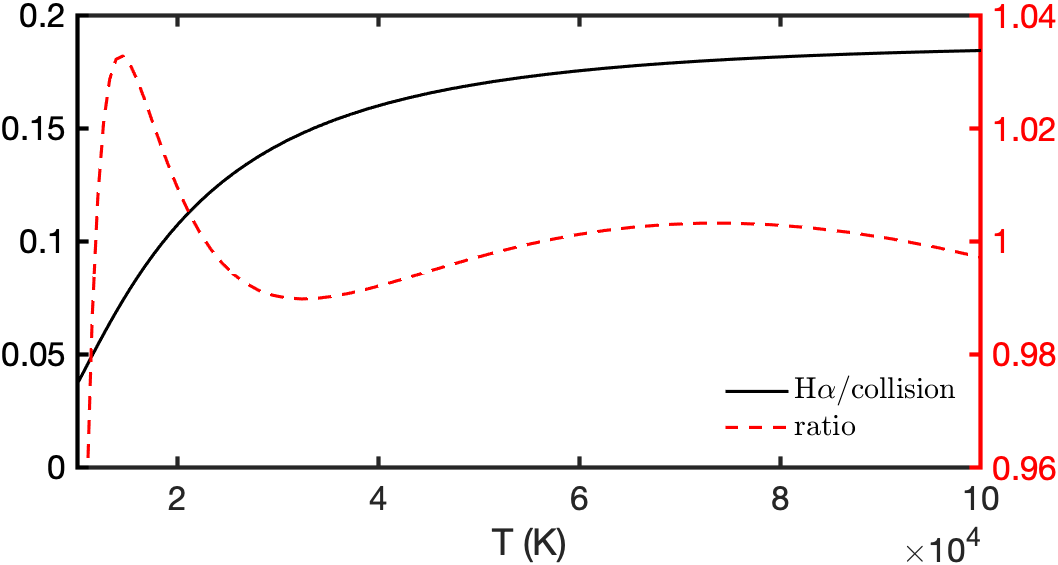}  %plot_Hcooling_warm.m
  \caption{(Left axis) The fraction, $f_{{\rm H}\alpha}$, of electron
  collisions with an \HI\ atom that result in emission of an H$\alpha$
  photon as a function of temperature.  A polynomial model for
  $y_{{\rm H}\alpha}$ is described in Equation~\ref{eq:f_Halpha}.
  The ratio of this model to calculations is displayed by the dashed
  red line.}
 \label{fig:Halpha_model_fit}
\end{figure}

In Figure~\ref{fig:Warmcooling_photon} we plot the line production 
efficiency\footnote{The fraction of photon (e.g. H$\alpha$, Ly$\alpha$) 
emitted per collision. Each two-photon emission is regarded as one event.}.  
Consistent with the collisional coefficients displayed in 
Figure~\ref{fig:Anderson_q} we see that Ly$\alpha$ has the highest 
efficiency, approximately 2/3, followed by 2-photon emission, $1/3$. 
H$\alpha$ is quite weak, even when measured by photons emitted.  
H$\alpha$ emission requires excitation to the $n=3$ level, whereas 
two-photon and Ly$\alpha$ emission are obtained by excitation to 
$n=2$ (and cascade from higher states).  However, H$\alpha$ has a 
major advantage ---it can be observed with existing ground-based 
observatories.  For this reason, we provide a fitting formula for  
$y_{\rm H\alpha}$, the fraction of H$\alpha$ photons per ionization,
 \begin{equation}
    f_{\rm H\alpha}=\sum_{k=0}^{2}a_k z^k \; ,
\label{eq:f_Halpha}
 \end{equation}
where $z = \log T_4$, as before. The model fit for H$\alpha$ is shown 
in Figure~\ref{fig:Halpha_model_fit}, and the values for the model
coefficients are given in Table~\ref{tab:photon_efficiency}, as well
as those for Ly$\alpha$ and two-photon emission.  

%  \vspace{2cm} 

\begin{deluxetable}{lrrr}[hbt]
\tablecaption{Efficiency of photon per collision}
\label{tab:photon_efficiency}
\tablewidth{0pt}
\tablehead{
\colhead{phot}&
\colhead{$a_0$} &\colhead{$a_1$} & \colhead{$a_2$}}
\startdata
H$\alpha$ & $0.031$ & $0.302$ & $-0.149$ \\
2$\gamma$ & $0.377$ & $-0.095$ \\
Ly$\alpha$ & $0.623$ & $0.095$ \\
\enddata
 \tablecomments{ Electron collisions with hydrogen atoms produce
Ly$\alpha$, two-photon continuum (2$\gamma$), H$\alpha$, and other
lines.  For each of these categories, the efficiency of photon
production, $f$, depends on temperature and is fitted to 
a model given by Equation~\ref{eq:f_Halpha}. The model fits are
accurate to 2\%. } 
\end{deluxetable}

%    \vspace{2cm} 

\section{Electron Collisional Ionization}
 \label{sec:Collisional_Ionization}
 
\begin{figure}[htbp]    	%Scholz_Black.m
 \plotone{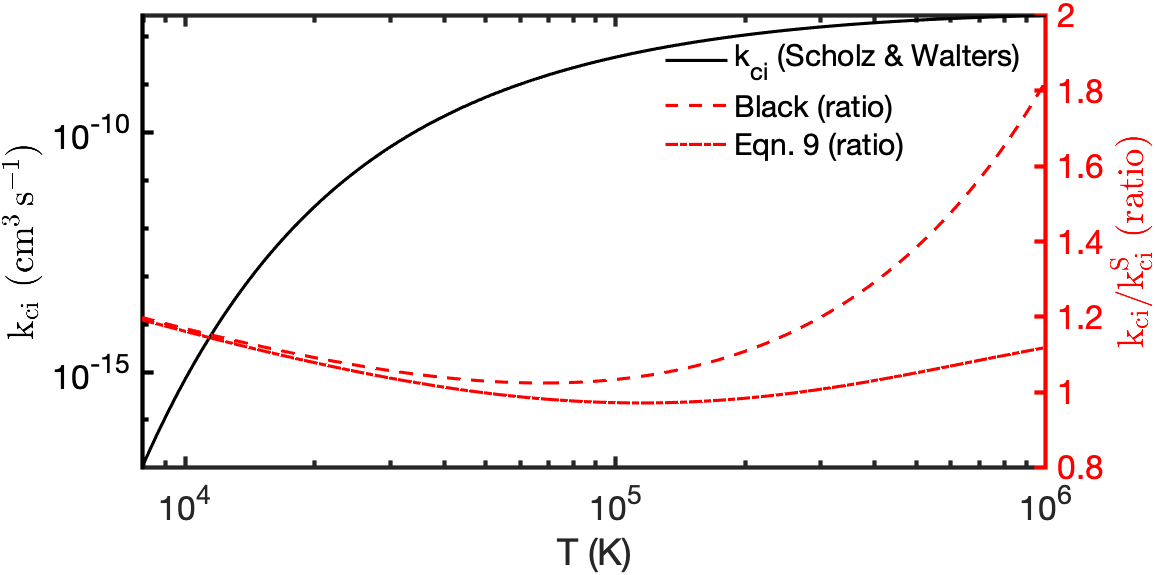}
  \caption{\small (Left axis): The run of collisional ionization
  coefficient (black line) of \citet{sw91}.  (Right axis): In red,
  we plot the ratio of the ionization rate coefficients of 
  \citet{B81} from Equation~\ref{eq:kci_Black} and our handy formula
  (Equation~\ref{eq:kci_Shull}) to that of Scholz \&\ Walters
  (1991) labeled $k_{ci}^{S}$. }
      \label{fig:H_excitation_Scholz_Black}
\end{figure}

The collisional ionization rate coefficient is derived 
as the integral,  
\begin{equation*}
  k_{ci}(T) = \int_{I_{\rm H}}^\infty \sigma_{ci}(E)v \, f(E)\, dE \; , 
 \end{equation*}
where $f(E)$ is the Maxwellian energy distribution, $I_{\rm H} =
13.598$~eV is the ionization energy of hydrogen, and $\sigma_{ci}(E)$
is the collisional ionization cross section as a function of electron
energy in the center-of-mass frame, $E=\sfrac{1}{2}\mu v^2$ with
$\mu=m_em_{\rm H}/(m_e+m_{\rm H})\approx m_e$.  Here, $m_e$ and
$m_{\rm H}$ are the mass of the electron and hydrogen atom,
respectively.  The collisional ionization rate coefficient can be
sensibly written as
 \begin{equation}
  k_{ci}(T) = A(T)\exp (- I_{\rm H}/k_B T)  \; .
\label{eq:kci}
 \end{equation}
\citet{B81} provided an approximate form for $k_{ci}T)$, based on
ionization cross sections tabulated by \citet{L67},
 \begin{equation}
     k_{ci}(T) = 5.85\times10^{-11}T^{\sfrac{1}{2}}
       \exp (-I_{\rm H}/k_B T) \, {\rm cm^{3}\,s^{-1}} \; .
   \label{eq:kci_Black}
 \end{equation}
This expression is consistent with the approximation, $\sigma_{ci}
\propto (1 - I_{\rm H}/E)$, quoted in \citet{D11}, but
only valid at low collision energies 
($I_{\rm H} \leq E \leq 3I_{\rm H}$).  As shown by \citet{L67}, 
the high-energy behavior is $\sigma_{ci} \propto \ln E/E$. 
\citet{sw91} provided a better approximation (for $10^4$~K to 
$2\times10^5$~K) using a sixth-order polynomial,
 \begin{equation}
  A(T) = \exp\Big(\sum_{i=0}^6 a_i y^i\Big)\ {\rm cm^3\,s^{-1}}
   \label{eq:kci_Scholz}
 \end{equation}
where $y= \ln T$.
A comparison (Figure~\ref{fig:H_excitation_Scholz_Black}) between
Equation~\ref{eq:kci_Black} (\citealt{B81} fit) and the more accurate
Scholz \&\ Walters fit (Equation~\ref{eq:kci_Scholz}) shows that
the former breaks down at high temperatures. We offer a modified 
formula with the correct asymptotic behavior ($k_BT > I_{\rm H}$) 
as used in the shock models of \citet{sm79},
 \begin{equation}
  k_{ci}(T)= \frac{5.85\times
  10^{-11}T^{\sfrac{1}{2}}}{\big[1+0.1k_BT/I_{\rm H}\big]} \exp
  \Big( -\frac{I_{\rm H}}{k_BT}\Big)\, {\rm cm^{3}\,s^{-1}}  \;  .
   \label{eq:kci_Shull}
 \end{equation}
As can be seen from Figure~\ref{fig:H_excitation_Scholz_Black},
this modified formula provides a good fit at both low and high
temperatures.  For quick estimates we use 
Equation~\ref{eq:kci_Shull}, but the polynomial formulation of 
\citet{sw91} is preferred when precision is needed (e.g., in 
numerical integration of differential equations).  The resulting
ionization power loss per unit volume is $n_en_{\rm HI}\Lambda_{ci}$ where  $\Lambda_{ci}=k_{ci}I_{\rm H}$ is the 
collisional ionization energy loss coefficient.

\section{The Hydrogen Cooling Curve}
 \label{sec:HydrogenCoolingCurve}

The goal in this section is to construct a cooling curve for ``warm"
($T\lesssim 10^5\,$K) hydrogen plasma. In \S\ref{sec:LineEmission}
we formulated the cooling coefficient due to line cooling, while in
\S\ref{sec:Collisional_Ionization} we presented the same for
ionization losses. Here, we summarize the cooling coefficients for
radiative recombination (free-bound) and free-free losses. Armed thus,
we formulate the cooling curve for hydrogen in the temperature range
$10^4\,$K to $10^5\,$K.

The kinetic energy of recombining electrons is a loss to the thermal
pool.  The model fits for $\alpha_k$  where $k=1, A, B$ (corresponding
to recombinations to $n=1s$ level, case~A and case~B) can be found in
Table~\ref{tab:Hummer_recomb_fit} (\S\ref{sec:KineticEnergy}).  The
radiative recombination power loss per unit volume is 
$n_en_p\alpha\Lambda_{\rm rr}$ where $\alpha$ is either case~A or case~B, 
as appropriate. Here, the radiative recombination energy rate coefficient 
$\Lambda_{\rm rr}=\alpha\langle E_{\rm rr}\rangle$ where 
$\langle E_{\rm rr}\rangle$ is the mean thermal energy lost by electron
upon recombination. Following \cite{D11} we let 
$\langle E_{\rm rr}\rangle = f_{\rm rr}k_BT$.

The  free-free emission rate per unit volume is 
$n_en_p\Lambda_{\rm ff}$, where $\Lambda_{\rm ff}$ is the free-free 
emissivity. The free-free power per electron is $n_p\Lambda_{\rm ff}$.  
The mean time for an electron to recombine is $(n_p\alpha)^{-1}$.  
Thus, the free-free energy lost up to the point of recombination is 
$\Lambda_{\rm ff}/\alpha$, which we equate to $f_{\rm ff}k_BT$.   
The combined recombination and free-free cooling rate coefficient is then
 \begin{equation}
	\Lambda_{\rm rf} =\Lambda_{\rm rr}+\Lambda_{\rm ff}=\alpha
	f_{\rm rf}(T)k_BT.  \label{eq:Lambda_ff} 
 \end{equation}
where $f_{\rm rf}=f_{\rm rr}+f_{\rm ff}$.  The run of $f_{\rm rf}(T)$
with temperature is displayed in Figure~\ref{fig:f_tot_AB}
(\S\ref{sec:KineticEnergy}), and the model fits are presented in
Table~\ref{tab:Hummer_recomb_fit} (\S\ref{sec:KineticEnergy}).

We now have all the elements to formulate the cooling rate per unit
volume, $\mathcal{C}(T)$, expressed as a negative value (for energy
losses):
 \begin{equation}
  \mathcal{C}(T)=-n_en_{\rm HI} \big[ \Lambda_{\rm HI}(T)
	+k_{ci}(T)I_{\rm H} \big] - n_en_p\Lambda_{\rm rf}  \; .
   \label{eq:CoolingCurve}
 \end{equation}
The three RHS terms are given by Equations~\ref{eq:Lambda_HI_formal},
\ref{eq:kci}, and \ref{eq:Lambda_ff}, respectively.

\section{Low-Velocity Shocks: A Simple Cooling Model}
 \label{sec:SimpleCoolingModel}

The investigation of time-dependent cooling of gas heated to $T
\approx 10^5$\,K is a classic endeavor, constituting the Ph.\,D.\
thesis topics of Michael Jura (\citealt{jd72}) and Minas Kafatos
(\citealt{K73}).  The motivation in the 1970s seems to have been
ambient gas heated by an FUV shock-breakout pulse. Separately,
\citet{ds78} investigated the production of Ly$\alpha$ from SN
shocks, and \cite{ss79} computed UV emission from SNRs in primeval
galaxies.

In unmagnetized plasma, the post-shock temperature of an adiabatic
shock is given by
 \begin{equation}
   T_s = \frac{2(\gamma-1)}{(\gamma+1)^2}\frac{\mu v_s^2}{k_B} =
   (1.12\times 10^5~{\rm K}) \frac{\mu}{m_{\rm H}}\Big(\frac{v_s}{70\,
   {\rm km\,s^{-1}}}\Big)^2\ \;  .
 \label{eq:T_s} 
\end{equation}
Here, $\mu$ is the mean mass per particle and $\gamma$ is the ratio
of specific heats at constant pressure and  constant volume; $\gamma
= 5/3$ for mono-atomic gas. If $y$ is the number density of helium
relative to that of hydrogen, the mean molecular mass for H$^0$ and
He$^0$ is $\mu = [(1+4y)/(1+y)] m_{\rm H}=1.23m_{\rm H}$ for
$y=0.0819$ \citep{2020A&A...641A...6P}.  For H$^+$ and He$^0$,
$\mu=0.64m_{\rm H}$.  For H$^+$ and He$^+$, $\mu=0.61m_{\rm H}$, and
for H$^+$ and He$^{+2}$, $\mu=0.59m_{\rm H}$.

Three timescales come into play for post-shocked gas: $\tau_r$, the
recombination timescale; $\tau_{ci}$, the collisional ionization
timescale; and $\tau_c$, the cooling timescale.  For gas around
$10^5\,$K, we have  $\tau_{ci} \ll \tau_r$.  With this inequality,
the cooling gas does not obey the conditions for collisional
ionization equilibrium.  Thus, it is often essential to undertake 
a full time-dependent calculation.

\subsection{Electron-Proton Equilibration}
 \label{sec:Electron-ProtonEquilibration}

At the collisionless shock front, the electrons and protons
receive similar amounts of random motion. Being more massive, the
protons acquire more energy and are initially much hotter
than the electrons.  The equilibration timescale for electrons to
be heated up to the temperature of the protons via electron-proton
encounters is approximately
 \begin{equation*}
  t_{\rm loss} = 14\ \Big(\frac{T}{10^5\,{\rm K}}\Big)^{3/2}
  \Big(\frac{\rm cm^{-3}}{n_e}\Big) \Big(\frac{25}{{\rm ln}\Lambda}
  \Big)\ {\rm yr}  \; ,
 \end{equation*}
where ${\rm ln}\Lambda$ is the Coulomb logarithmic factor accounting
for distant encounters (\citealt{S78}; Chapter 2). The current view
\citep{lrm96,glr07} is that plasma instabilities and electromagnetic
waves drive electron-proton equilibration faster than two-body
interactions.  We assume that equipartition occurs before collisional
ionization sets in (\S\ref{sec:CollisionalIonization}).

\vspace{1cm}

\subsection{Collisional Ionization}
 \label{sec:CollisionalIonization}
 
The rate equation for the number density of electrons is 
 \begin{equation}
  \frac{dn_e}{dt} = n_en_{\rm HI} k_{ci}(T)-n_e n_p\alpha(T) \; ,
   \label{eq:CI_1}
 \end{equation}
where $\alpha=\alpha_A, \alpha_B$ as needed.
Given our assumpton of hydrogen plasma, the number density
of protons, $n_p=n_e$.
As noted in \S\ref{sec:Algebra}
a solution to this equation at cosnatnt $T$ is
 \begin{equation*}
   x(t)^{-1} = x_0^{-1}\exp(-t/\tau_{ci}) +
  x_{eq}^{-1}\Big[1-\exp(-t/\tau_{ci})\Big]  \; .
 \end{equation*}
Here, $x_0=x(t=0)$, $\tau_r\equiv (\alpha n_{\rm H})^{-1}$ and
$\tau_{ci}\equiv (k_{ci}n_{\rm H})^{-1}$ are the characteristic time
scales for recombination and collisional ionization, respectively,
and
 \begin{equation}
  x_{eq} (T)=
  \frac{k_{ci}(T)}{\alpha+k_{ci}(T)}=\frac{\tau_r}{\tau_r+\tau_{ci}}
   \label{eq:xeq}  \ .
 \end{equation}
The lowest probable value for the ionization fraction in a realistic
diffuse atomic medium, before heating commences, is $x_0\approx
2\times 10^{-4}$.  In this case the electrons come from stellar FUV
photoionization of C, S, Mg, Si, Fe and other trace metals.  The timescale
for electron  ionization to reach fraction $x$ is  $\tau_{ci}
\ln (x/x_0)$.

\begin{figure}[htbp]		%plot_Recombination_Ionization.m
 \plotone{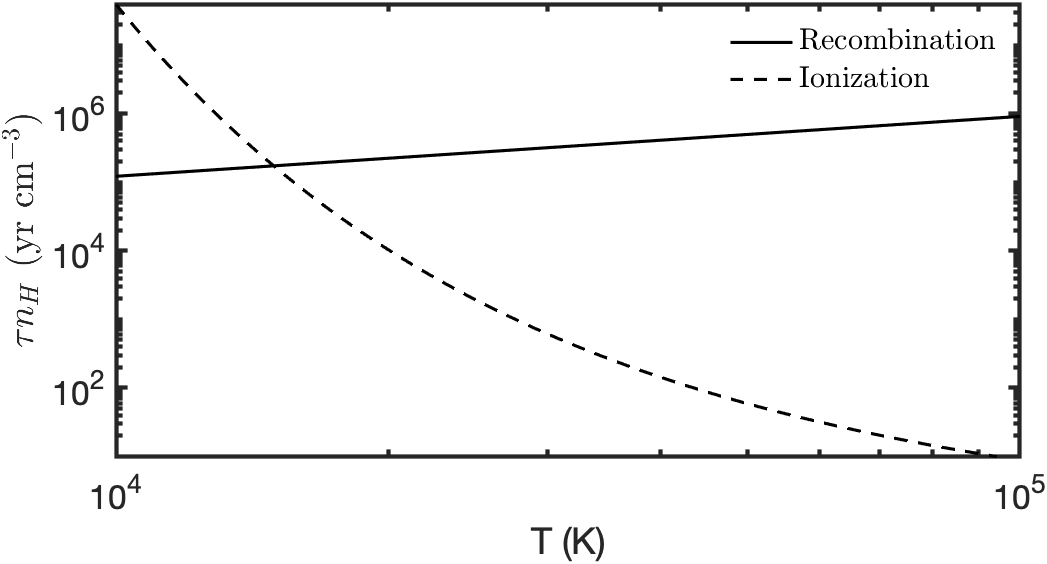}
  \caption{\small Ionization time ($\tau_{ci}$; see
  Equation~\ref{eq:kci_Shull}) and recombination time ($\tau_r$;
  case B) as a function of temperature.  The two timescales cross
  at about 15,000\,K, at which point the ionization fraction would
  be 50\%, if collisional ionization equilibrium were to hold (see
  Equation~\ref{eq:xeq}).}
 \label{fig:tau_ci_r}
\end{figure}

\subsection{Recombination}
 \label{sec:Recombination}

As can be seen from Figure~\ref{fig:tau_ci_r}, collisional ionization
is a strong function of temperature.  At late times, when the plasma
has cooled, collisional ionization can be ignored and
Equation~\ref{eq:CI_1} simplifies to $dx/dt=-x^2/\tau_r$, with the
solution
 \begin{equation*}
  \frac{1}{x}-\frac{1}{x_0} = \frac{t}{\tau_r}.
 \end{equation*}
The ionization fraction decreases from $x_0$ to $x_0/m$ on a timescale
of $t=(m-1)\tau_r/x_0$.  The run of recombination time scale as a
function of temperature is shown in Figure~\ref{fig:tau_ci_r}.

\subsection{Basic Cooling and Recombining Framework}
 \label{sec:Cooling}

The path in the phase diagram of density, ionization fraction, and
temperature along which the gas cools depends on the cirucmstances.
For planar radiative shocks, the pressure behind the shock is $P_0
+ (3/4)\rho_0 v_0^2$, rising to $P_0 + \rho_0v_0^2$ downstream when
$\rho\gg \rho_0$.  Here, the pre-shock parameters have subscript
0. Thus, radiative shocks are good examples of cooling at nearly constant
pressure (``isobaric").  As the gas cools, its density rises to
maintain the pressure.  A second possibility is
cooling at constant density
(``isochoric"). The decrease in temperature, following cooling,
leads to lower pressure.  Pressure changes are conveyed at the speed
of sound, $c_s$. The time scale for adiabatic sound waves to cross
a nebula of length $L$ is $\tau_a=L/c_s$.  Isochoric cooling will
take place when the cooling time is short, $\tau_c\ll \tau_a$.

The first law of thermodynamics states that any gain in the internal
energy ($U$) of the system is due to increase in internal heat and
work done: $dU=dQ-PdV$. For mono-atomic gas, the internal energy
of the nebula per unit volume is $U=(3/2)nk_BT$, while the pressure
is given by Boyle's law $P=nk_BT$.  Here, $N=nV$ is the total number
of particles in the nebula whose volume is $V$. Let $N_{\rm H}=n_{\rm
H}V$ be the total number of hydrogen nuclei. Ionization can produce
changes in $N$, whereas $N_{\rm H}$ is fixed.

The three physical parameters  governing the cooling hydrogen plasma
are $n_e$, $T$ and $n_{\rm H}$. We have two differential equations,
one for ionization balance ($n_e$; Equation~\ref{eq:CI_1}) and one
for for energy loss ($T$; discussed below).  A third differential
equation follows from the assumed framework: $dP/dt=0$ (isobaric
cooling) or $dV/dt=0$ (isochoric cooling).

For an isochoric system, no work is done by or on the nebula.
Adopting case~B framework the energy balance equations becomes
 \begin{equation*}
	q\frac{d}{dt}(nk_BT)=-n_en_{\rm HI}\big[\Lambda_{\rm HI}
	  +k_{ci}I_{\rm H}\big] -n_e^2\alpha_B f_{\rm rf} k_B T  \;  ,
 \end{equation*}
where $q=3/2$ and the RHS gives the total cooling rate.  Note that
$n_{\rm H}$ remains constant, whereas $n = n_{\rm H} + n_e$ varies
as the ionization fraction changes.  The ionization-recombination
equation (Equation~\ref{eq:CI_1}) can be restated
 \begin{equation*}
  \frac{dn}{dt} = n_en_{\rm HI}k_{ci} - n_e^2\alpha_B.
 \end{equation*}
We combine the above two equations to obtain
 \begin{equation*}
  \begin{split}
 qnk_B\frac{dT}{dt} =-n_en_{\rm HI}\big[\Lambda_{\rm HI}+k_{ci}I_{\rm
 H}+k_{ci}qk_BT\big] \\
     -n_e^2\alpha_B \big(f_{\rm rf}-q\big)k_B T \; , 
   \end{split}
 \end{equation*}
which we deliberately recast as
 \begin{equation}
  \begin{split}
 qnk_B\frac{dT}{dt} =-n_en_{\rm HI}\big[\Lambda_{\rm HI}
     +k_{ci}I_{\rm H}\big] - n_e^2\alpha_B f_{\rm rf}k_B T \\
  + qk_BT\big[n_e^2\alpha_B-n_en_{\rm HI}k_{ci}\big] \; . 
  \end{split}
  \label{eq:qnkB}
 \end{equation}
In this formulation, the meaning of Equation~\ref{eq:qnkB} is clear.
The LHS arises from the loss of internal energy. The first term on
the RHS represents energy loss from \HI\ collisional line excitation
($\Lambda_{\rm HI}$) and collisional ionization (loss of $I_{\rm
H}$  per collision).  The loss of kinetic energy per recombination
(including the free-free radiation up until the recombination
event) is given by the second term.  The final term accounts for
losses/gains to the thermal pool of electrons during recombination
and ionization.  For plasma in collisional ionization equilibrium
this term vanishes, as expected.

For isobaric cooling, the pressure, $P=(n_{\rm H}+n_e)k_BT$ is
fixed. In this case, we compute $n_e$ and $T$ and then deduce $n_{\rm
H}$ through the pressure equation. As the nebula cools, the ambient
gas, in order to maintain the pressure, $P_a$, does work on
the nebula by compressing the nebula.   The work done by the 
medium on the nebula is $P dV$. However, since $P$ is constant,
$d(PV)=P_adV$.  The internal energy of the nebula is then the
enthalpy, $HV$ where $H=U+P_a$.  Going forward, we will drop the
subscript to $P$.  It is this store of enthalpy that powers the
nebular cooling, $\mathcal{C}V$. Since  $(U+P)V= (5/2) Nk_BT$ we
see that Equation~\ref{eq:qnkB} still applies but with $q=5/2$.

\begin{comment}

For numerical integration, Equation~\ref{eq:qnkB} can be restated
as
 \begin{equation*}
  \frac{dT}{dt} = -\frac{x(1-x)}{(1+x)}\frac{T}{\tau_c} -
  \frac{x^2}{(1+x)} \bigg[\frac{f_r}{q}-1\bigg]\frac{T}{\tau_r}
 \end{equation*}
where $\tau_c= qk_BT/[n_{\rm H}(\Lambda_{\rm HI}+k_{ci}I_{\rm
H}+k_{ci}qk_BT)]$ is the cooling time scale, and $\tau_r=(n_{\rm
H}\alpha_B)^{-1}$ is the recombination time scale
(\S\ref{sec:CollisionalIonization}).

\end{comment}

\subsection{A shock heated nebula}
 \label{sec:Cooling_H_nebula}
 
Consider a nebula composed of hydrogen which has been shocked heated
to, say, $T_s=10^5\,$K. The electron-proton equilibration timescale
will be shorter than $14 n_e^{-1}$~yr (see
\S\ref{sec:Electron-ProtonEquilibration}).  The collisional ionization
time, $t_{\rm ci}$, is short, $10n_{\rm H}^{-1}$ yr at $T=10^5\,$K,
rising to $60n_{\rm H}^{-1}$~yr at 50,000\,K.  Thus, even if the
pre-shocked gas has minimal ionization ($x_0\approx 2\times 10^{-4}$)
it will take a time, $\tau_{\rm ci} \ln(x/x_0) = 7.8t{\rm ci}$ for
the ionization fraction to reach $x = 0.5$.  The initial losses are
large, owing to both collisional ionization and collisional excitation
by the newly liberated electrons and subsequent radiation.  An
exception is if the pre-shocked gas is pre-ionized.  If the shock
is strong, pre-ionization (H$^+$ and He$^+$) will be achieved, which
diminishes the hydrogen Ly$\alpha$ emission. (There will still be
cooling from He~II Ly$\alpha$ $\lambda304$ and lines from metal
ions.) As the gas cools, recombination becomes more efficient.  Once
the gas reaches $10^4\,$K, cooling by forbidden lines of metals
will occur.  The gas will eventually settle down at $T_1\approx
5000$--$8000$\,K, the temperature of the stable WNM phase (see
\citealt{ht03,ksc+03,pkc+18,msg+18}).

Ignoring ``metals", the mean particle mass is $\mu=m_{\rm
H}(1+4y)/(1+y+x_0)$.  The shock velocity and $\mu$ determine the
post-shock temperature, $T_s$ through Equation~\ref{eq:T_s}.  Recall
that, in our simplified model, the losses from the shocked nebula
are only those associated with hydrogen (line radiation, ionization,
free-bound, and free-free). In particular, while we include helium
in computing the reduced mass, we do not include losses due to
helium.  In short, we treat helium as a silent and inactive partner.
The energy per H-nucleus and associated electron is $E_0=q k_B
T_s(1+x_0)$. The end state is when hydrogen has largely recombined
and thus the energy per H-atom is $E_1=q k_B T_1$.

For our fiducial temperature of $10^5\,$K, we have $E_0\approx [12.9,
21.5](1+x_0)$\,eV energy per H atom for $q=3/2, 5/2$.  Thus, on
simple grounds, we can see that low-velocity shocks will not
significantly affect the ionization of the incoming particles.
More precise radiative shock models show that fast shocks,
$v_s > 110\,{\rm km\,s^{-1}}$, can pre-ionize (H$^+$, He$^+$) the 
incoming medium \citep{sm79,R79,ds96}.

\begin{figure}[htbp]
 \plotone{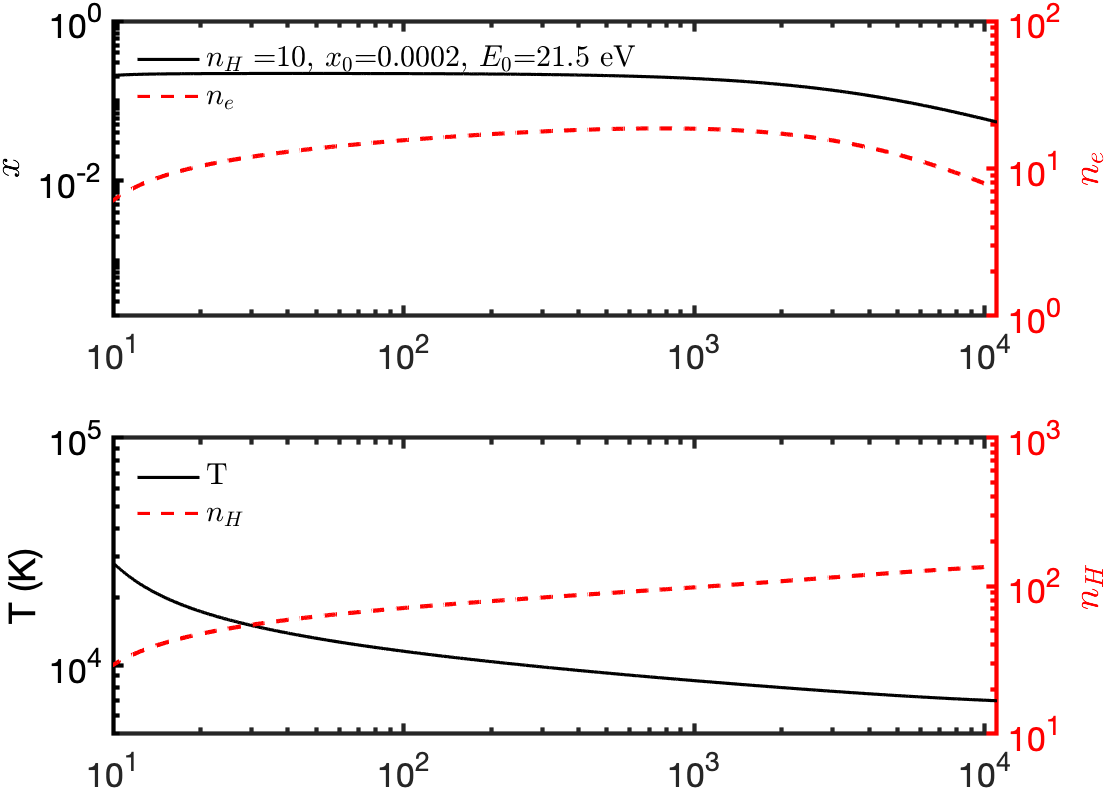}\smallskip
  \plotone{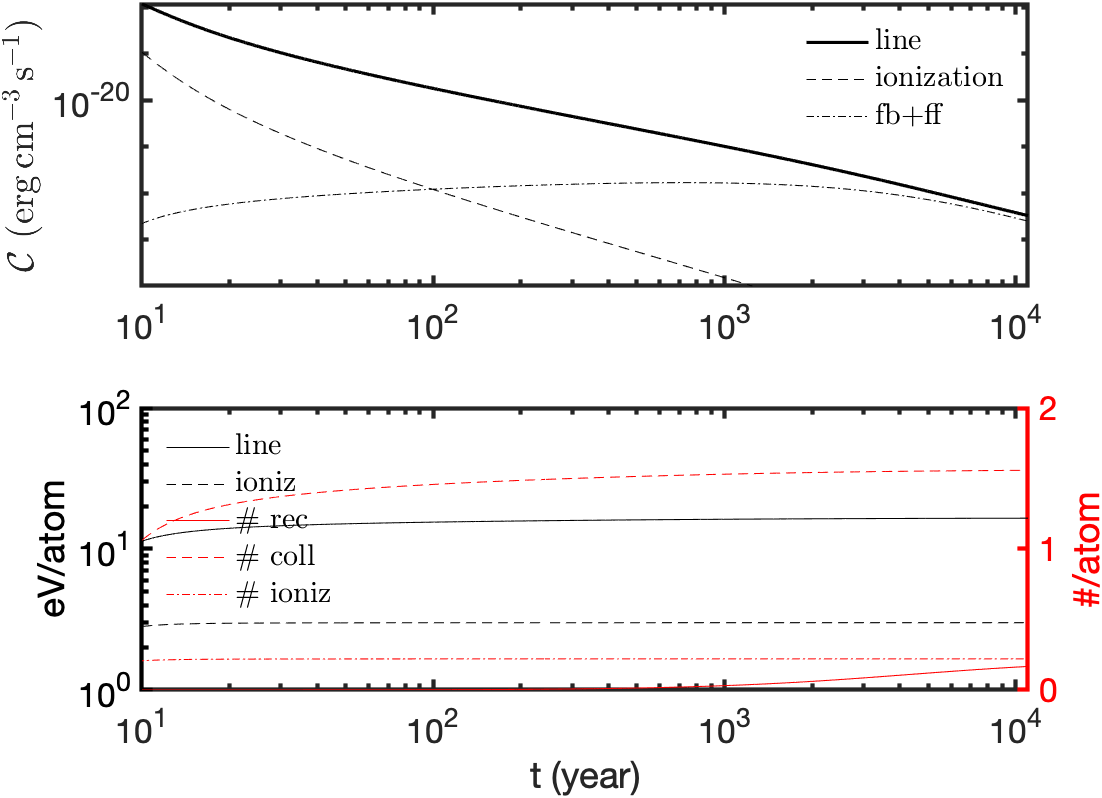} 
   \caption{\small Run of temperature and density of a pure hydrogen
   plasma suddenly heated to $10^5\,$K and subsequently cooling
   down via an isobaric process.  The hydrogen density, $n_H$ and
   ionization, $x_0$ at $t=0$ are given in the legend in the top-most
   panel.}
 \label{fig:run_1}
\end{figure}

\begin{figure}[htbp]
 \plotone{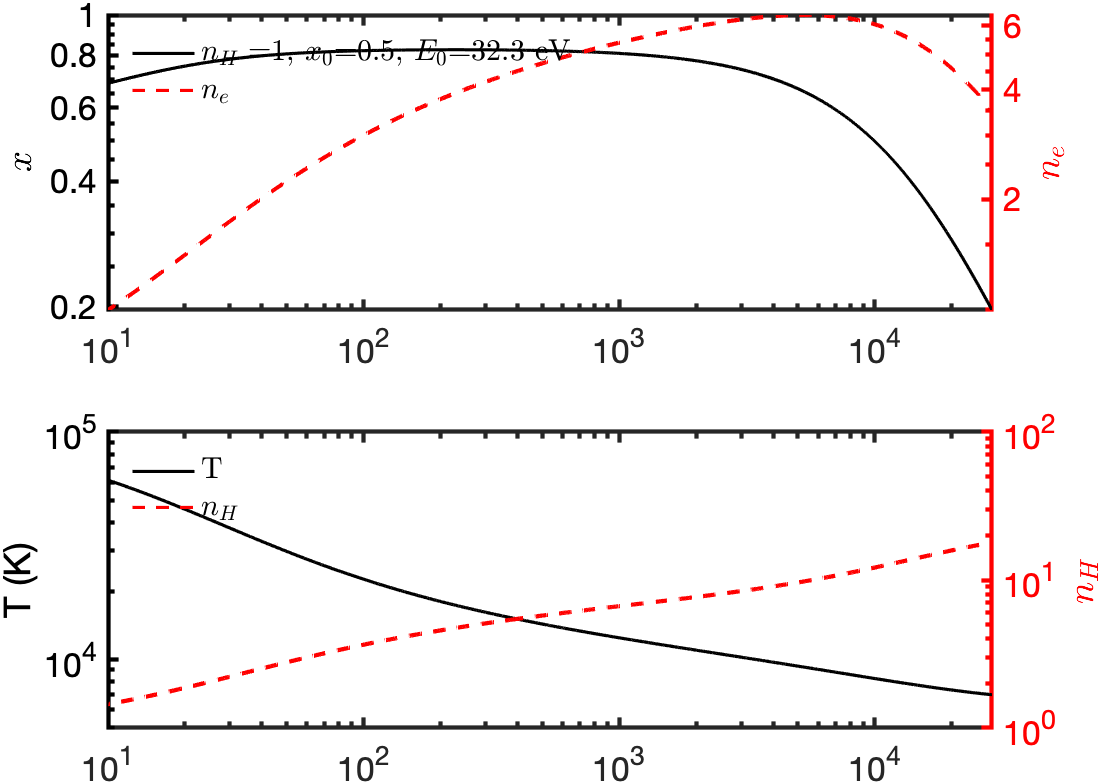}\smallskip
  \plotone{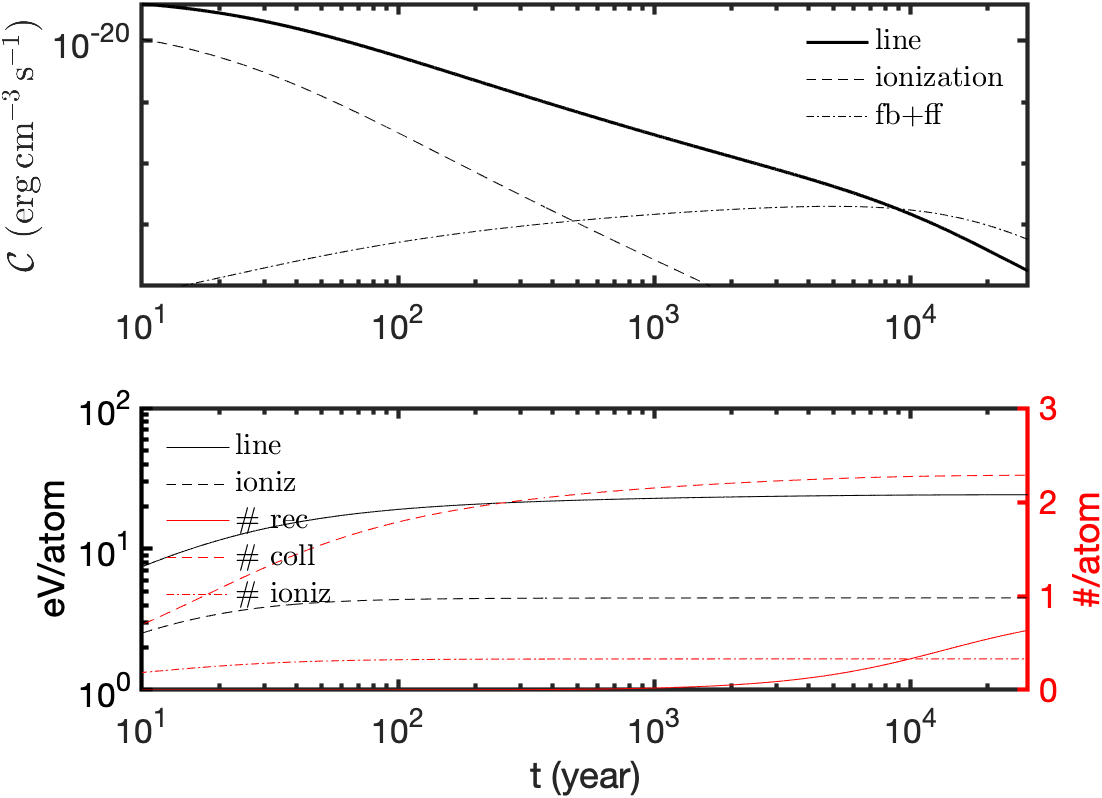} 
   \caption{\small
      See caption to Figure~\ref{fig:run_1}. }
 \label{fig:run_2}
\end{figure}

Example runs are shown in Figures~\ref{fig:run_1} and \ref{fig:run_2}.
In addition to the run of physical quantities ($T$, $x$, $n_{\rm
H}$) we also plot the total number of recombinations per H nucleus,
 \begin{equation*}
   N_r=\int \frac{1}{n_{\rm H}} n_{\rm H}^2x^2\alpha_B(T)dt  \; ,
 \end{equation*}
and the total number of collisions per atom
 \begin{equation*}
	N_c= \int n_{\rm H}x(1-x)\sum_{k=1}^{14}q_{k}(T)dt \; .
 \end{equation*}

\vspace{2cm}

\section{Conclusion \&\ Prospects}
\label{sec:ConclusionProspects}

Low velocity shocks with velocities near 70\,km\,s$^{-1}$ abound in
our Galaxy. Some descend from higher velocity shocks (e.g., supernova
remnants) while others start at low velocity (e.g., stellar
bow shocks, high velocity cloud shocks). These shocks do not have
strong pre-cursor ionization fronts, and as such the post-shocked
gas is partially neutral. Such shocks cool primarily through
Ly$\alpha$, two-photon continuum, H$\alpha$, and metal emission lines.
Ly$\alpha$ is the brightest line, although resonant scattering traps 
usually traps these photons within the plasma, resulting in absorption 
by dust grains. H$\alpha$ is weak but has the great advantage of being 
observable from the ground.

Two-photon continuum emission is about 50\% of Ly$\alpha$ emission 
(see Figure~\ref{fig:Warmcooling_photon}). It is several times
brighter than H$\alpha$, even when one compares photon fluxes
rather than energy fluxes. Fortunately, two-photon emission can be
observed with space-based observatories. Furthermore, the two-photon 
continuum has a distinct FUV/NUV ratio.  In fact, {\it GALEX} FUV and 
NUV imagery has led to the recent discovery of large middle-aged supernova
remnants \citep{fdw+21} and exotic shocked stellar bow shocks with
angular scales of hundreds of degrees \citep{bba+20}.  The Ultraviolet
Explorer (UVEX) is a  NASA Explorer mission currently under development 
\citep{khg+22}.  Amongst other goals, UVEX aims to undertake FUV and NUV 
imaging of the entire sky with higher sensitivity and finer spatial 
resolution, relative to {\it GALEX}.
The aforementioned successes with {\it GALEX} imagery show great promise
of identifying and studying low velocity shocks in a future 
all-sky survey with UVEX.  

With this motivation, and using the best available atomic physics data
and atomic calculations, we computed the collisional and cooling
coefficients for warm hydrogen ($T\lesssim 10^5\,$K).  The primary
application of our results is in computing two-photon continuum from 
bow shocks and old supernova remnants.  We allow for pre-ionization 
by keeping the ionization fraction of the pre-shocked gas as a free 
parameter that can be set to values computed from more sophisticated 
shock models ({\it ibid}; \citealt{ds96}).  Our expectation is that 
the accurate H-cooling developed here can be incorporated into 
time-dependent models (e.g., \citealt{gs07}).

For completeness, we discuss two-photon emission from photoionized
gas (e.g., H~II regions, the Warm Ionized Medium).  Draine (2011;
Table 14.2)\nocite{D11} provides the recombination coefficient to
the 2s level, $\alpha_{\rm 2s}$ and the recombination coefficient
for H$\alpha$ emission. From this we find ratios
 \begin{equation}
  \frac{\alpha_{\rm 2s}}{\alpha_B} \approx 0.328 T_4^{0.115},\ 
   \frac{\alpha_{\rm H\alpha}}{\alpha_B} \approx 0.450 T_4^{-0.11} \; . 
 \end{equation}
Thus, at typical temperatures of photoionized gas, recombination 
process 
result in similar diffuse emission for two-photon continuum and
H$\alpha$. However, while H$\alpha$ emission 
is concentrated in a narrow line, the two-photon continuum is
distributed over the FUV band. Compensating for this effect, the FUV 
sky is incredibly dark relative to the optical band (see \citealt{K22} 
for detailed analysis of the FUV background).

In the Galactic plane and at low latitudes, two-photon emission will
be attenuated by dust in the intervening neutral ISM and contaminated
by reflected light from dust grains.  In practice, this means that 
the use of two-photon continuum as a diagnostic will be restricted to high Galactic
latitudes and will require careful modeling of reflected light. However, 
the early success with {\it GALEX} promises rich returns from the all-sky 
survey in both the FUV and NUV planned with UVEX.

\acknowledgements
We thank Nikolaus Zen Prusinski, California Institute of Technology,
for help with CHIANTI, a collaborative project involving George
Mason University, the University of Michigan (USA), University of
Cambridge (UK) and NASA Goddard Space Flight Center (USA).

\bibliography{bibHCooling}{}
\bibliographystyle{aasjournal}

\appendix

\section{Comparison to other cooling models}
 \label{sec:OtherCoolingModels}

In this section we compare our cooling curve, $\Lambda_{\rm HI}(T)$,
discussed in \S\ref{sec:HydrogenCoolingCurve}, to some notable
published curves.

\begin{figure}[htbp] %Scholz_Anderson.m
 \centering
  \includegraphics[width=3in]{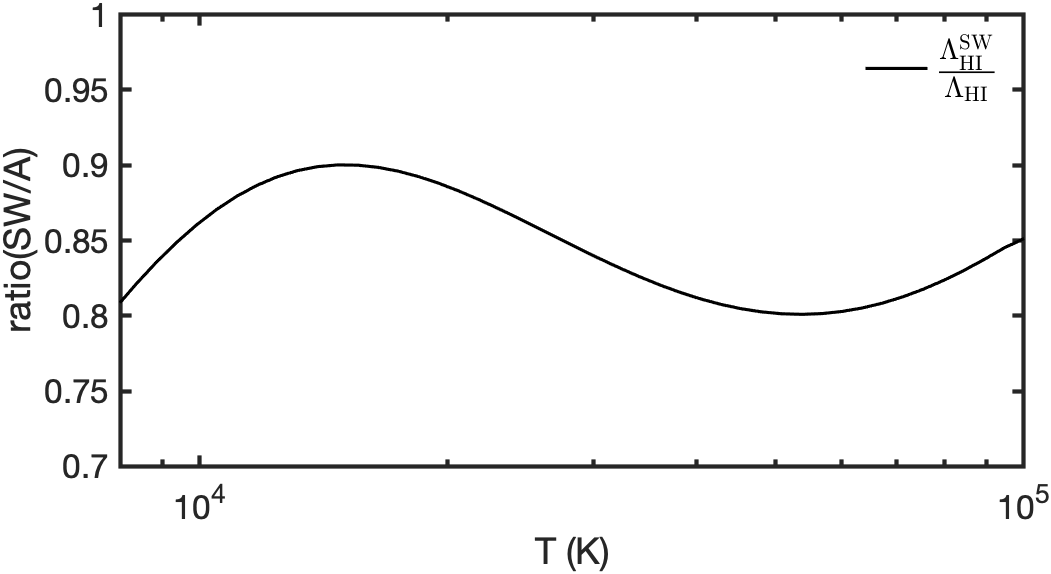}\ \ 
   \includegraphics[width=3in]{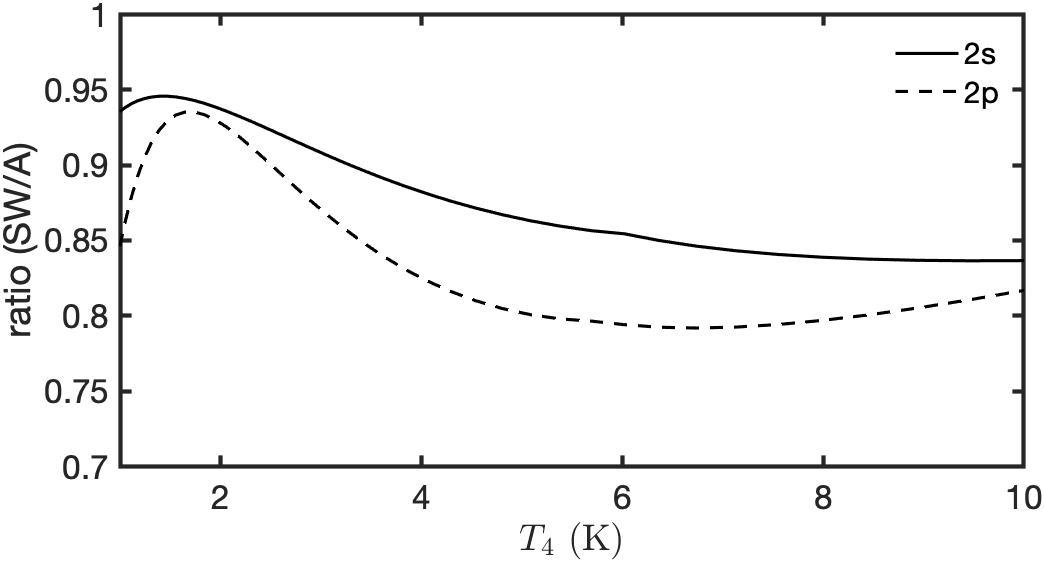} 
  \caption{\small (Left) Comparison of hydrogen line-cooling rate
  from \citet{sw91}, $\Lambda_{\rm HI}^{\rm SW}(T)$ given in
  Equation~\ref{eq:Lambda^prime}, with our $\Lambda_{\rm HI}(T)$
  (Equation~\ref{eq:Lambda_HI_formal}).  The ordinate is the ratio
  $\Lambda_{\rm HI}^{\rm SW}(T)/\Lambda_{\rm HI}(T)$.  (Right).
  Comparison of the collisional line excitation coefficients of
  {\it 1s}$\rightarrow${\it 2s} and {\it 1s}$\rightarrow${\it 2p} 
  transitions between \cite{sw91} (``SW"; see Equation~\ref{eq:q1s_SW}, 
  and the corresponding equation of \cite{abb+02} (``A").  The 
  ordinate is the ratio $q_{1s\rightarrow f}$ from \cite{sw91} 
  to that of \cite{abb+02}.}
 \label{fig:Scholz_Anderson}
\end{figure}

\subsection{Scholz \& Walters (1991)}
 \label{sec:ScholzWalters}
 
\cite{sw91} provide collisional ionization coefficient
(Equation~\ref{eq:kci_Scholz}) and total cooling  (line and ionization
energy losses) coefficients. The latter is given by the following
polynomial model:
 \begin{equation}
	\Lambda_{\rm H}^{\rm SW}(T)= 10^{-20}\times \exp\Big(\sum_{i=0}^5
	d_i y^i - \frac{T_{12}}{T}\Big) \ {\rm erg\,cm^3\,s^{-1}}
		\label{eq:Lambda^prime}
 \end{equation}
where $T_{12}= (3/4)T_R$ with $k_BT_R=I_{\rm H}$ and $y={\rm ln} T$.
We subtracted losses to ionization, $\Lambda_{ci}=k_{ci}I_{\rm H}$,
to obtain $\Lambda_{\rm HI}^{\rm S}(T)$. In
Figure~\ref{fig:Scholz_Anderson} we compare our $\Lambda_{\rm HI}(T)$
(Equation~\ref{eq:Lambda_HI_formal}) to $\Lambda_{\rm HI}^{\rm
S}(T)$.  The Scholz-Walters curve is systematically lower by 15\%.
After correction for this scale factor, the two curves agree to
within $\pm 5\%$. Separately, \citet{sw91} codified the collisional
coefficients for excitation from the ({\it 1s}) ground state of hydrogen
to $f =$ {\it 2s} and {\it 2p} levels
 \begin{equation}
  q_{1s\rightarrow f}(T)=\Gamma_{1s\rightarrow f}(T) \exp (-T_{12}/T)
   \label{eq:q1s_SW}
 \end{equation}
where $\Gamma_{1s\rightarrow f}$ is formulated as
 \begin{equation*}
  \Gamma_{1s\rightarrow f}(T) = \exp\Big(\sum_{i=0}^5 b_i y^i\Big)\,
  {\rm cm^3\,s^{-1}}  \; .
 \end{equation*}
In Figure~\ref{fig:Scholz_Anderson} we compare these two coefficients
with our own coefficients (\S\ref{sec:LineEmission}).  As with the
total line emission cooling, the {\it 2s} and {\it 2p} collisional 
rate coefficients are discrepant by the same scale factor.

\subsection{Spitzer (1978)}

\begin{figure}[htbp] %CHIANTI_Anderson.m Spitzer_Anderson.m
 \centering
  \includegraphics[height=1.7in]{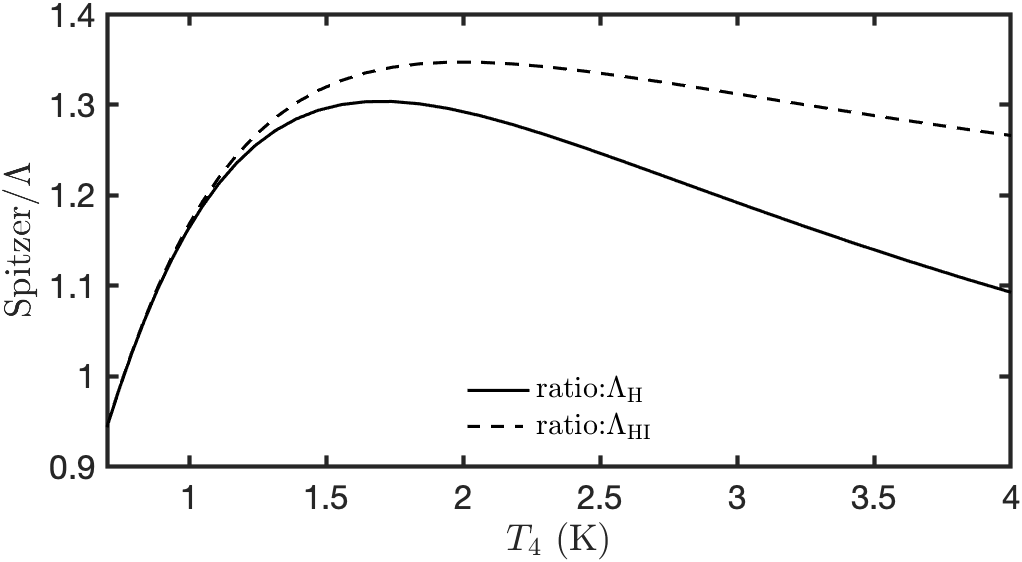}\qquad
   \includegraphics[height=1.7in]{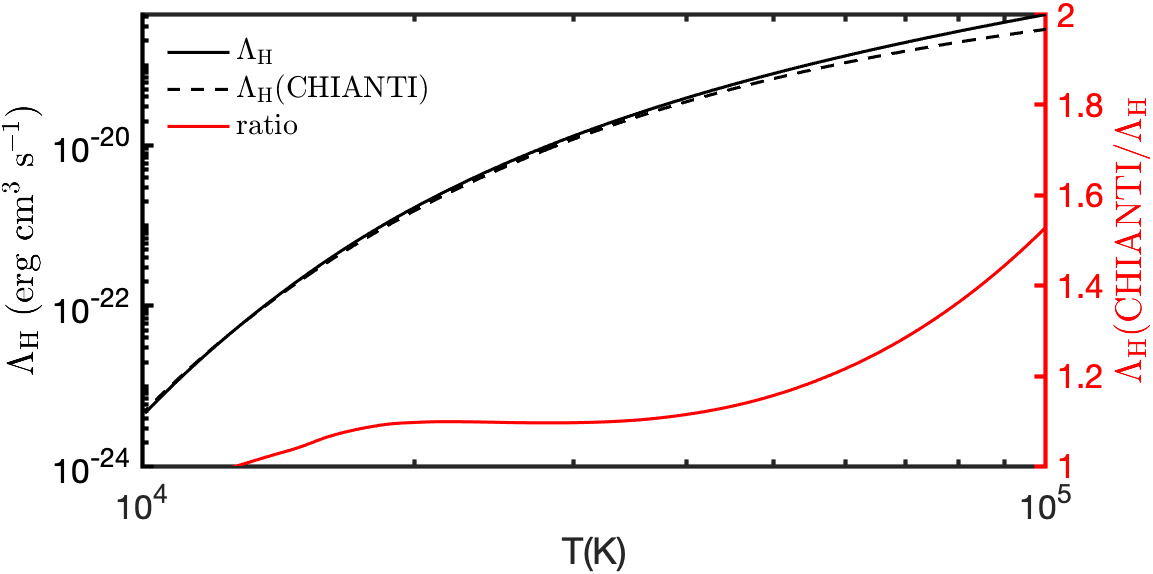}
    \caption{\small (Left). The ratio of Spitzer's H-line cooling
    function to our computations of $\Lambda_{\rm HI}$ (line cooling)
    and $\Lambda_{\rm H}$ (including collisional ionization cooling)
    as a function of temperature.  (Right). Comparison of the total
    cooling curve (line and ionization losses) between our cooling
    curve and that from CHIANTI.}
 \label{fig:Spitzer_CHIANTI}
\end{figure}

A classical formula for the volumetric cooling rate of hydrogen
plasma is given in \citet{S78}, $\mathcal{C}=n_e n_{\rm H}\Lambda_{\rm
Spitzer}$, where
 \begin{eqnarray}
   \Lambda_{\rm Spitzer}(T) &=&7.3\times 10^{-19}\exp(-T_{12}/T)\,{\rm
   erg\,cm^3\,s^{-1}}.
	\label{eq:Spitzer}
 \end{eqnarray}
We are aware that our cooling coefficient formulation expresses the line
cooling rate per unit volume as $n_en_{\rm HI}\Lambda_{\rm HI}$.
However, there is little difference between our formulation and
Spitzer's formulation when the ionization fraction is small. This
fit to $\Lambda_{\rm HI}(T)$ was claimed to be accurate to 3\% over
the range 4,000\,K to 12,000\,K.  This formula does not include
losses from collisional ionization. Spitzer's fit was actually made
to \HI\ cooling rates taken from Table~2 of \citet{dm72} which were
based on \HI\ excitation cross sections assembled by \citet{gt70} from 
theoretical calculations in the 1960s.  Those atomic calculations are 
superseded by more recent work used in this paper \citep{abb+02}.  
In the left panel of Figure~\ref{fig:Spitzer_CHIANTI} we display the 
ratio of our cooling curves, 
$\Lambda_{\rm HI}$ and $\Lambda_{\rm H} =\Lambda_{\rm HI}+k_{ci}I_{\rm H}$ 
to the cooling rate of Spitzer (Equation~\ref{eq:Spitzer}).  Our fit is 
tuned to be accurate over the range $10^4\,{\rm K}$ to $10^5\,{\rm K}$.  
Even bearing this in mind, it appears that Spitzer's formula over-estimates 
line cooling at temperatures above $10^4\,K$ by about 25\%.

\subsection{Collisional Ionization Equilibrium: CHIANTI }

We conclude this section by briefly discussing hydrogen plasma which
is in collisional ionization equilibrium (CIE; electron ionization
balanced by radiative recombination). The volumetric cooling rate
in CIE is given by $\mathcal{C}=x_{\rm eq}(1-x_{\rm eq})n_{\rm
H}^2\Lambda_{\rm CIE}$ where $x_{\rm eq}$ is given by
Equation~\ref{eq:xeq}.  It is mainly dominated by line cooling.
CHIANTI \citep{dlm+97,ddy+21} is a major resource for astronomers
working on collisionally excited gas, especially hot plasma which
are in CIE. CHIANTI returns $x_{\rm eq}$ and cooling function
$\mathcal{C}\equiv n_en_{\rm H}\Gamma$ (under assumption of case~A).
Thus, $\Gamma=(1-x_{\rm eq})\Lambda_{\rm H}$.

\begin{figure}[htbp]   %CIE.m
 \centering
  \includegraphics[width=4.5in]{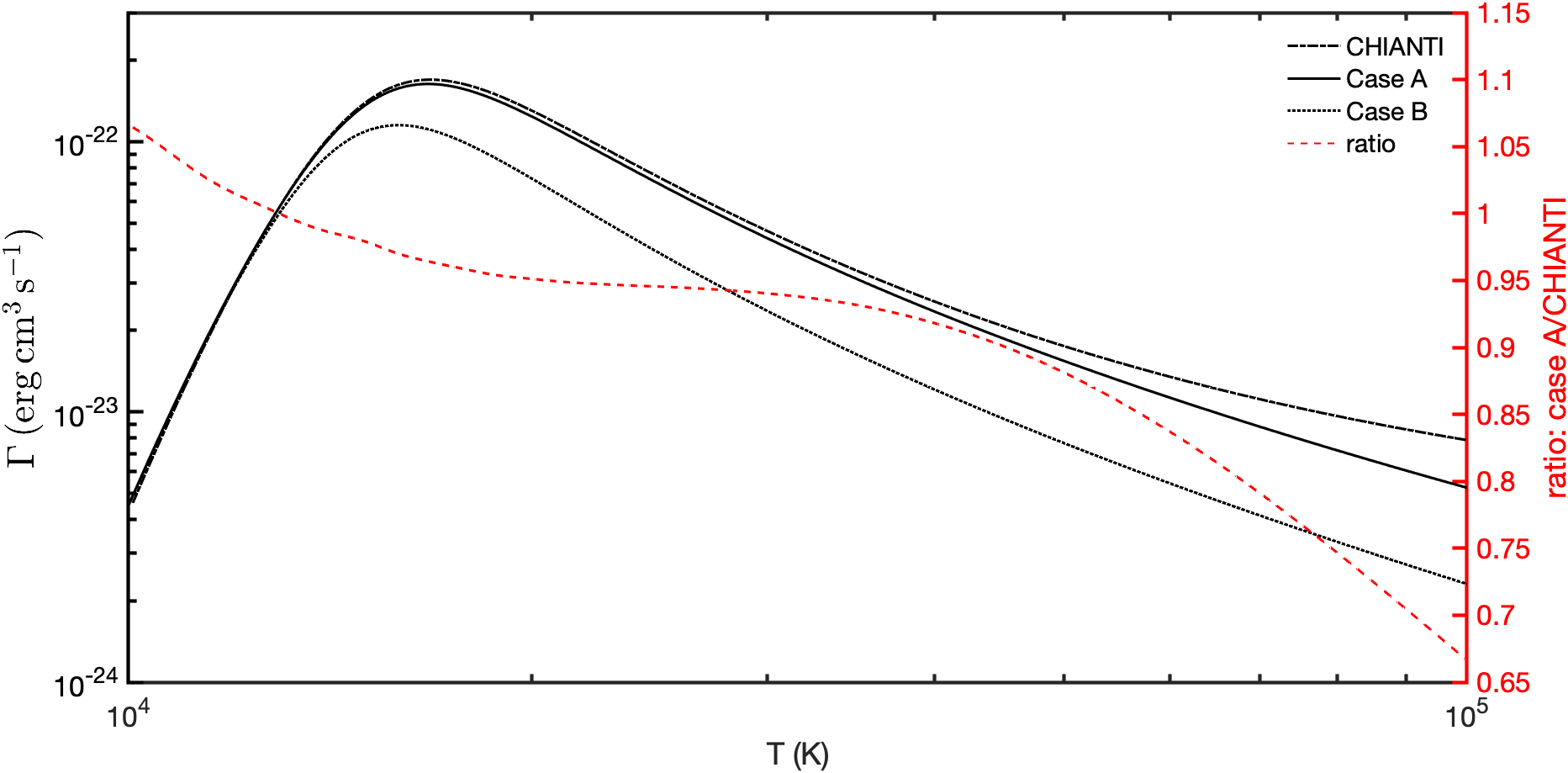} 
   \caption{\small (Left): The run of $\Gamma$ as obtained from
   CHIANTI and the calculations reported here (case A, CIE) with
   temperature $T$.  Here, the $\Gamma_{\rm CIE}$ is defined as
   follows: $\mathcal{C}=n_en_{\rm H}\Gamma_{\rm CIE}$ where it is
   understood that $\mathcal{C}$ and $n_e$ reflect CIE conditions.
   (Right): The ratio of $\Gamma$ (this work) to that reported from
   CHIANTI. }
 \label{fig:CIE_gamma} 
\end{figure}

\begin{figure}[htbp] 	%CIE.m
 \centering
  \includegraphics[width=3in]{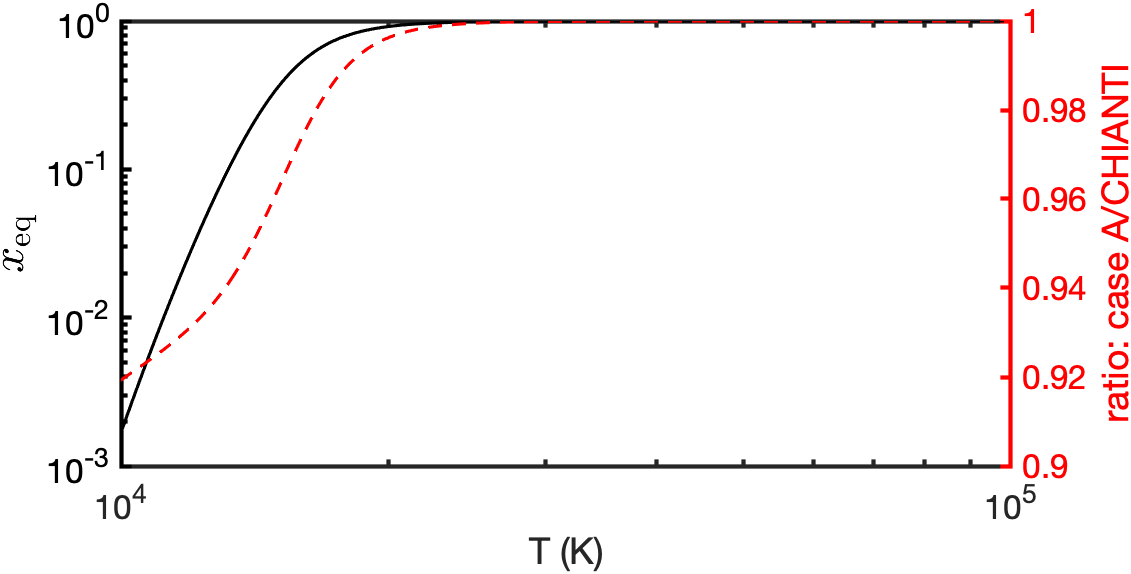}
   \caption{\small (Left) The run of $x_{\rm eq}$ under CIE conditions
   as a function of temperature, $T$.  (Right) The ratio of $x_{\rm
   eq}$ (this paper) to that provided by CHIANTI.  }
 \label{fig:CIE_xe} 
\end{figure}

The run of $\Gamma$ from  with temperature is displayed in
Figure~\ref{fig:CIE_gamma}. It is evident that $\Gamma$ from CHIANTI
is brighter relative to the calculations reported here at higher
temperatures.  There is a small difference As can be seen from
Figure~\ref{fig:CIE_xe} our  $x_{\rm eq}$ agrees with that used in
CHIANTI for $T>2\times 10^4\,$K.

%SRK: APNMWD
%Hiormi: HDOPQI

\section{Losses due to recombination, free-free emission and
ionization}
 \label{sec:KineticEnergy}

\begin{figure}[htbp] 
 \centering
 \includegraphics[width=3in]{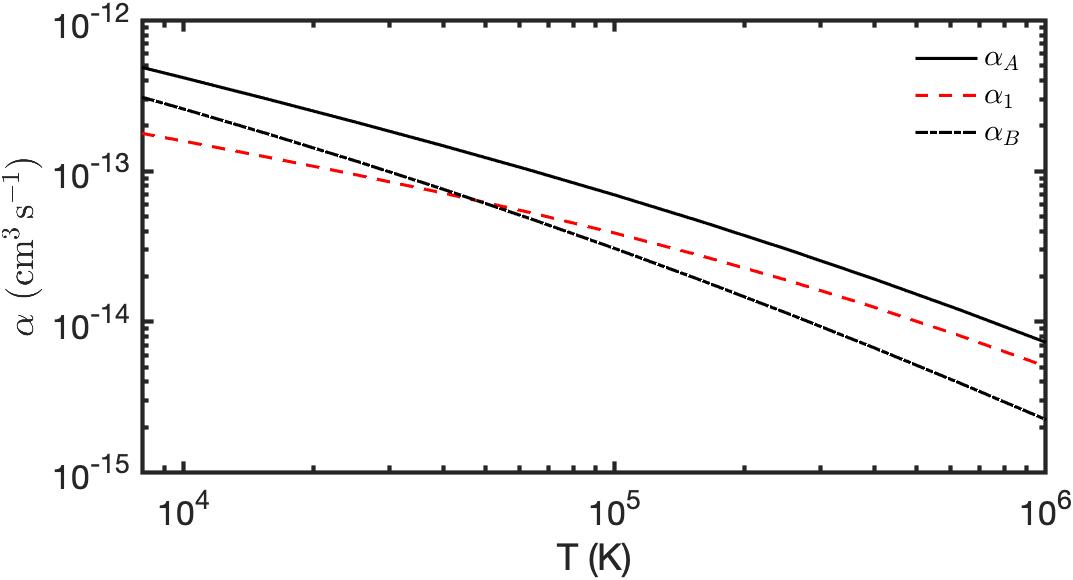}
   \caption{\small Run of H-recombination coefficients (from
   \citealt{H94}).  }
 \label{fig:Hummer_alpha} 
\end{figure}

Standard textbooks (e.g., \citealt{D11}) provide fitting formulae
for hydrogen (recombination, free-bound losses) tuned for study of
H~II regions. Here, we present fitting formula in the temperature
range interest to this paper: $10^4$\,K to $10^5$\,K. Our starting
point is \cite{H94} who, over an impressive range of 10\,K to
$10^7$\,K, present recombinations rate coefficients, $\alpha_i$
where $i=1$ (recombination to $n=1$) and $i=A, B$ for case~A and
case~B; see Figure~\ref{fig:Hummer_alpha}.

\begin{figure}[htbp] 
 \centering
  \includegraphics[width=3in]{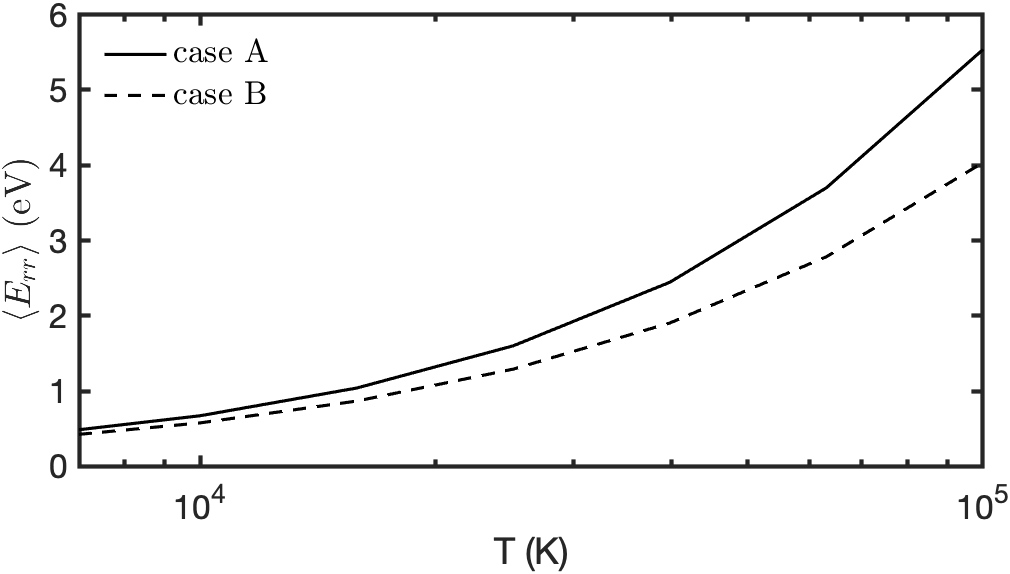}\qquad\qquad
   \includegraphics[width=3in]{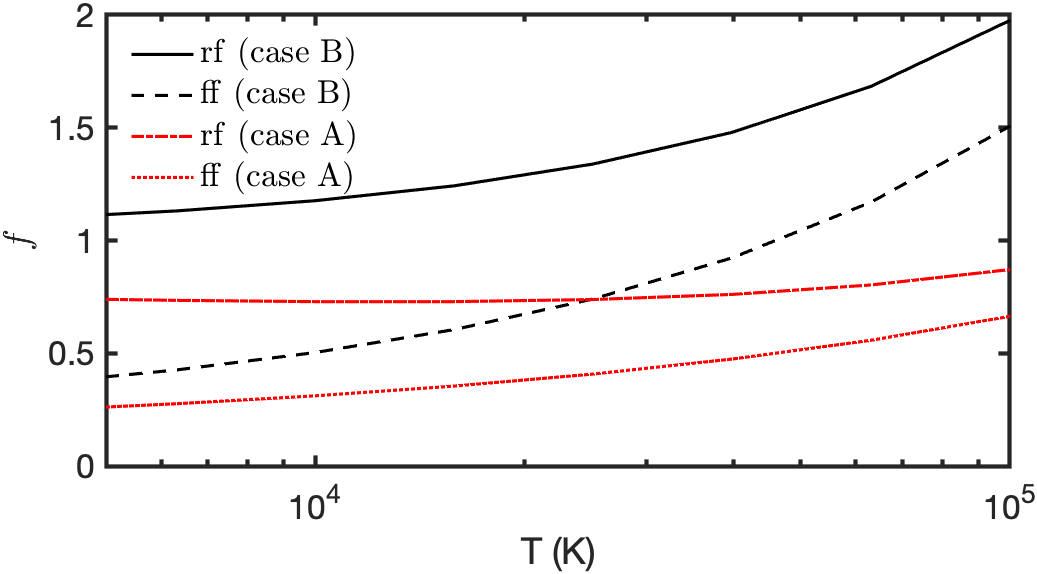}
  \caption{\small (Left): The run of the average kinetic energy of the
  recombining electron, $\langle E_{\rm rr}\rangle$,  as a function
  of temperature.  (Right): The run of $f$ as a function of temperature
  ($T$).  The superscripts denote case A or case B. The subscript is 
  ``rr" (free-bound radiative recombination), ``ff"
  (free-free emission), and $f_{\rm rf}=f_{\rm ff}+f_{\rm rr}$.  }
 \label{fig:f_tot_AB}
\end{figure}

Separately, \citet{H94} also tabulate the kinetic energy loss due
to recombination and free-free emission.  In Figure~\ref{fig:f_tot_AB}
we plot the mean kinetic energy, $\langle E_{\rm rr}$ versus $T$).
Note that $\langle E_{\rm rr}\rangle=f_{\rm rr}k_BT$.   The resulting fits are presented in
Table~\ref{tab:Hummer_recomb_fit} and are accurate to one percent.

\begin{deluxetable}{lrrr}[hbtp]
\tablecaption{Fits to $\alpha$ and $f$}
 \label{tab:Hummer_recomb_fit}
\tablewidth{0pt}
\tablehead{
\colhead{qty} & \colhead{$A$} & \colhead{$n$} & \colhead{$b$}}
\startdata
$\alpha_1$ & $1.58\times 10^{-13}$ & $-0.518$ & $-0.039$\\
$\alpha_A$ & $4.16\times 10^{-13}$ & $-0.708$ & $-0.030$\\
$\alpha_B$ & $2.58\times 10^{-13}$ & $-0.822$ & $-0.045$\\
\hline
$f^{\rm A}_{\rm rr}$ & 0.784 &  $-0.042$  &   $-0.020$ \\
$f^{\rm B}_{\rm rr}$ & 0.672 &  $-0.109$ &  $-0.021$\\
$f^{\rm A}_{\rm rf}$ & 1.09   &   $0.035$ &   $0.019$\\
$f^{\rm B}_{\rm rf}$ & 1.17   &  $0.087$  &   $0.061$\\
\enddata
 \tablecomments{``qty" is fitted to the model of the form 
 ${\rm qty}=AT_4^{n+b{\rm ln} T_4}$ where the temperature range 
 is $5\times 10^3\,{\rm K} < T < 2\times 10^5\,{\rm K}$; here,
 $T_4=T/(10^4\,{\rm K})$. See text for definition of subscripts.
 The superscripts stand for case~A or case~B.  The unit for $\alpha$
 is ${\rm cm^3\,s^{-1}}$ while that for $f$ is dimensionless.
 The fits are accurate to 1\%\ over the temperature range of interest.
 }
\end{deluxetable}

For free-free emission rate coefficient, $\Lambda_{\rm ff}$ we use
Equation~10.12 of \cite{D11} and from that derive $f_{\rm ff}(T)$
(see \S\ref{sec:HydrogenCoolingCurve} for definition of $f_{\rm rf}$).
The values of $f_{\rm rf}=f_{\rm rr}+f_{\rm ff}$ for case A 
and case B are displayed in Figure~\ref{fig:f_tot_AB} (left panel).
We assume a simple linear relation and obtain the following fits
for $f_{\rm rf}$:
 \begin{equation*}
   f_{\rm rf}=0.71+0.0154T_4\ ({\rm case~A}),\qquad
   f_{\rm rf} =1.10+0.0898T_4\ ({\rm case~B})\ .
 \end{equation*}
 
We conclude this section with a discussion of collisional ionization.
The energy of the electron, $E$, upon collision goes into ionizing
the atom and imparting kinetic energy to the newly liberated electron.
We use the low-energy approximation (Equation~\ref{eq:kci_Black})
for the collisional cross-section, $\sigma_{ci}\propto [1-(E/I_{\rm
H})]$ (valid for $I_{\rm H} \leq E \leq 3I_{\rm H}$) to find a mean
kinetic energy following ionization,
 \begin{eqnarray*}
  \langle E_k\rangle &=& \langle E-I_{\rm H}\rangle = \frac{\int_{I_{\rm
  H}}^\infty \sigma_{ci}(E)v f_E(E-I_{\rm H})dE}{k_{ci}(T)} \approx
  2k_BT  \; .
 \end{eqnarray*}
This formula becomes inaccurate at high temperature ($kT \gtrsim 3I_{\rm
H}$) where the adopted fit to $\sigma_{ci}$ breaks down ($\sigma_{ci}$
peaks and then falls off as $\ln E / E$). The energy $\langle E_k\rangle$ 
is shared between the colliding electron and the ionized electron.
This energy is not a loss since it is returned to the thermal pool.
However, over time,  the ionized electron will draw $qk_BT$ energy
($q=3/2$ in isochoric framework and $q=5/2$ in isobaric framework)
from the thermal pool. This is a genuine loss (see the discussion
following Equation~\ref{eq:qnkB}).

\section{Two solutions to Recombination-Ionization equation}
 \label{sec:Algebra}

Equation~\ref{eq:CI_1} can be written as
 \begin{eqnarray*}
  \frac{dx}{dt} &=&  \frac{x}{\tau_{ci}} - \frac{x^2}{\tau_h}
	\label{eq:dxdt_appendix}   \;  , 
 \end{eqnarray*}
where $\tau_h^{-1}=\tau_{ci}^{-1}+\tau_{r}^{-1}$ is the harmonic
mean of the two timescales.  The equilibrium value for ionization
is obtained by setting the LHS to zero, $x_{eq}=\tau_h/\tau_{ci}$.
The above equation can be re-arranged to yield
 \begin{eqnarray*}
  \frac{dx^\prime}{x^\prime[1-x^\prime]}&=&\frac{dt}{\tau_{ci}} \; ,
 \end{eqnarray*}
where $x^\prime=x/x_{eq}$. This equation can be integrated using
the method of  partial fractions to yield
 \begin{equation*}
  {\rm ln}\Big\vert\frac{x^\prime}{1-x^\prime}\Big\vert=\frac{t}{\tau_{ci}}
  + {\rm const.}
 \end{equation*}
At $t=0$, $x^\prime=x_0/x_{eq}\equiv x_0^\prime$ and thus
 \begin{equation*}
  \frac{t}{\tau_{ci}} ={\rm
     ln}\Bigg\vert\frac{x^\prime(1-x_0^\prime)}{x_0^\prime(1-x^\prime)}\Bigg\vert.
 \end{equation*}
This equation is useful to compute the time-scale to achieve a particular
level of ionization.  As $t\rightarrow\infty$, $x\rightarrow x_{eq}$,
as expected. 
Alternatively, we apply  the transformation, $u=1/x$:
 \begin{equation*}
  \Big[\frac{du}{dt} + \frac{u}{\tau_{ci}}\Big] = \frac{1}{\tau_h}.
 \end{equation*}
Multiply both sides by $e^{t/\tau_{ci}}$ to obtain
 \begin{equation*} \Big[ e^{t/\tau_{ci}}\frac{du}{dt} +
 \frac{u}{\tau_{ci}}e^{t/\tau_{ci}}\Big] = \frac{1}{\tau_h}e^{t/\tau_{ci}}.
 \end{equation*}
Because the LHS is the derivative of $ue^{t/\tau_{ci}}$, the
above equation can be readily integrated to yield
 \begin{equation*}
  u(t) = u_0e^{-t/\tau_{ci}}+ u_{eq}\Big[1-e^{-t/\tau_{ci}}\Big]
 \end{equation*}
where $u(t=0)=u_0=x_0^{-1}$ and $u_{eq}=x^{-1}_{eq}$.  As
$t\rightarrow\infty$, as expected, $u^{-1}\rightarrow x_{eq}$.

\end{document}